\documentclass[seceq,preprint]{ptptex}

\usepackage{amsmath,amsfonts,latexsym,amssymb}
\usepackage{verbatim,amsthm,graphicx}
\usepackage{wrapft}

\usepackage{amsmath}
\usepackage{amssymb}
\usepackage{mathrsfs}

\def\defscript{\mathscr}

\def\LL{{\defscript L}}

\def\ifempty#1{\def\tmpdata{#1}\ifx\tmpdata\empty }

\def\linebreak{\hfill\break}

\def\bra<#1|{\langle #1\rvert}
\def\ket|#1>{\lvert#1 \rangle}
\def\braket<#1|#2>{\langle #1|#2 \rangle}

\def\pfrac#1#2{\left(\frac{#1}{#2}\right)}
\def\const{\text{const}}
\def\otop#1{\hbox{$#1\kern-0.1em$\llap{\hbox{\raise1.7ex\hbox{$\scriptstyle\circ$}}}} }

\def\inpare#1{\left(#1\right)}
\def\bigpare(#1){\left(#1\right)}
\def\inrbra#1{\left\{ #1 \right\}}
\def\insbra#1{\left[ #1 \right]}

\def\bigbra[#1]{\left[ #1 \right]}

\def\h{\hat }

\def\b{\bar }

\def\tend{\rightarrow}
\def\then{\Rightarrow\quad}

\def\therefore{\mbox{\setbox0=\hbox{X}\hbox{$\ldotp$}\raise0.7\ht0\hbox{$\ldotp$}\hbox{$\ldotp$}} \quad }
\def\because{\mbox{\setbox0=\hbox{X}\raise0.7\ht0\hbox{$\ldotp$}\hbox{$\ldotp$}\raise0.7\ht0\hbox{$\ldotp$}}\kern0pt }

\def\SO{{\rm SO}}

\def\upin{\hbox{\setbox0=\hbox{$\cup$} \vrule width 0.05 \wd0 height \ht0 depth 0pt \kern - 0.5\wd0 \box0 }}
\def\Frac(#1/#2){\left(\frac{#1}{#2}\right)}

\def\sdprod{\mathrel{{\setbox0=\hbox{$\displaystyle\times$}\lower0.3\wd0\hbox{$\stackrel{\box0}{\scriptstyle\sim}$}}}}

\def\tosigma#1,{%
    \ifx\tmpindex\relax \def\tmpindex{#1} \let\next=\tosigma
    \else \ifnum\tmpindex=0 1 \else \sigma_\tmpindex \fi
          \ifx#1\relax  \let\next=\relax
          \else \otimes \let\next=\tosigma \def\tmpindex{#1} \fi
    \fi \next}
\def\tspb(#1){\let\tmpindex=\relax\tosigma#1,\relax,}
\def\Order#1{{\rm O}\!\left(#1\right)}

\def\diff#1#2{\frac{d #1}{d #2}}
\def\mdiff#1#2#3{\frac{d^{#1} #2}{d #3^{#1}}}

\def\pd{\partial}

\def\THB{{\mathbb T}}
\def\VHB{{\mathbb V}}
\def\SHB{{\mathbb S}}

\def\Eq#1{\begin{equation} #1 \end{equation}}

\def\Eqr#1{\begin{eqnarray} #1 \end{eqnarray}}

\def\Eqrsub#1{\begin{subequations}\Eqr{#1}\end{subequations}}
\def\Eqrsubl#1#2{\begin{subequations}
  \expandafter\ifx\csname Rlabel\endcsname \relax \label{#1}
  \else \Rlabel{#1} \fi \Eqr{#2}\end{subequations}}
\def\Bitm{\begin{itemize}}
\def\Eitm{\end{itemize}}
\def\Blist#1#2{\begin{list}{#1}{\parsep=0pt \itemsep=0pt%
  \listparindent=0pt #2}}
\def\Elist{\end{list}}
\long\def\ignore#1#2{\def\ignoreflag{#1}\long\def\tmptext{#2}
  \ifnum\ignoreflag>1 #2 \fi}

\def\THB{{\mathbb T}}
\def\VHB{{\mathbb V}}
\def\SHB{{\mathbb S}}

\notypesetlogo
\preprintnumber[3cm]{\begin{tabular}[t]{l}KEK-Cosmo-9\\KEK-TH-1248\end{tabular}}
\title{\large Accelerating a Black Hole in Higher Dimensions}

\author{Hideo {\sc Kodama}%
\footnote{E-mail: Hideo.Kodama@kek.jp}
}
\inst{Cosmophysics Group, IPNS, KEK and the Graduate University of Advanced Studies,
1-1 Oho, Tsukuba 305-0801, Japan
}

\abst{
Utilising the master equation with source for perturbations of the Schwarzschild-Tangherlini solution, we construct perturbative solutions representing a black hole accelerated by a string in higher dimensions. We show that such solutions can be uniquely determined by a single function representing the local tension of the string, under natural asymptotic and regularity conditions. We further study whether we can construct a localised braneworld black hole solution from such a solution by cutting off a region containing the string by a hypersurface and putting a vacuum brane on the boundary. We find that the solution corresponding to the string with constant tension does not allow such brane configuration when the bulk spacetime dimension is greater than four, in contrast to the four dimensional case. Further, we show that there exist infinitely many localised braneworld black hole solutions in the perturbative sense for four-dimensional bulk spacetime, if we allow non-uniform string tensions.
}

\begin{document}
\maketitle

\section{Introduction}

It has been shown that the Randall-Sundrum braneworld model with a single brane\cite{Randall.L&Sundrum1999b} may provide a viable model for the real world that can replace the conventional four-dimensional model. For example, the Robertson-Walker universe was implemented in that model reproducing the standard cosmological model for the late stage of our Universe\cite{Mukohyama.S2000,Mukohyama.S&Shiromizu&Maeda2000}. Further, it was shown that the behavior of cosmological perturbations of the brane in such an implementation is very close to that in the conventional four-dimensional model at least in the stage in which the cosmic expansion rate is smaller than the curvature of the bulk adS spacetime\cite{Koyama.K&Soda2000,Kodama.H&Ishibashi&Seto2000,Kodama.H2000A,Kobayashi.T&Tanaka2006}. 

In contrast to these cosmological aspects, the viability of the Randall-Sundrum model in the astrophysical problems is quite unclear. In particular, although this model was shown to reproduce the Newtonian gravity on large scales in the weak field limit\cite{Randall.L&Sundrum1999b,Garriga.J&Tanaka2000}, it is not certain whether its predictions on astrophysical phenomena associated with strong gravity are the same as or similar to those of the Einstein theory in the conventional four-dimensional framework.

The most important issue related to this is the existence and uniqueness of a static localised vacuum black hole solution corresponding to the Schwarzschild black hole solution in the four-dimensional Einstein theory. Here, by a localised black hole, we mean a black hole whose horizon has a compact spatial section, unlike the warped black string. No exact solution representing such a localised black hole has been found nor has been shown to exist exactly in five or higher dimensional models yet\cite{Chamblin.A&Hawking&Reall2000,Kanti.P&Tamvakis2002,Karasik.D&&2004,Karasik.D&&2004a,Dadhich.N&&2000,Casadio.A&Fabbri&Mazzacurati2002,Chamblin.A&&2001,Kofinas.G&Papantonopoulos&Zamarias2002,Kodama.H2002a,Casadio.R&Mazzacurati2003}. Further, although such solutions were numerically constructed in the case of small horizon sizes compared to the bulk adS curvature scale\cite{Kudoh.H&Tanaka&Nakamura2003}, no one has succeeded in constructing a static localised black hole solution whose horizon size is much larger than the bulk curvature scale even numerically\cite{Shiromizu.T&Shibata2000,Tanahashi.N&Tanaka2008}. Some are even predicting that such a localised black hole would not exist on the basis of the adS/CFT correspondence\cite{Tanaka.T2003,Tanaka.T2007A}. 

This situation suggests that there may not exist even a localised static braneworld black hole with a small mass in the exact sense. In the present paper, we develop a formulation to study this problem by a perturbative method. 

\begin{figure}[t]
\centerline{
\includegraphics*[height=5cm]{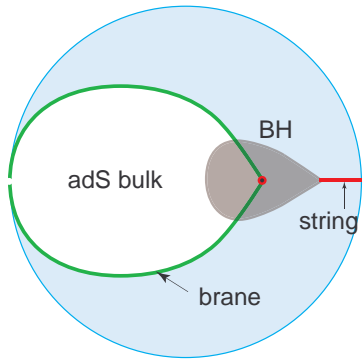}
}
\label{fig:Cmetric4D}
\caption{A localised black hole in the 4D braneworld model}
\end{figure}

The basic idea comes from observations of a static localised black hole solution in the braneworld model with four-dimensional bulk spacetime. As was first pointed out by Emparan, Horowitz and Myers\cite{Emparan.R&Horowitz&Myers2000,Emparan.R&Horowitz&Myers2000a}, such a solution can be constructed from the generalised C-metric representing an accelerated black hole in four-dimensional adS spacetime\cite{Plebanski.J&Demianski1976}. This C-metric has a conical singularity along one side of the symmetry axis passing through the black hole, which corresponds to a string with constant tension physically and provides acceleration for the black hole. The braneworld black hole solution can be constructed from this solution by cutting off a half of the spacetime by an appropriate hypersurface crossing the horizon and putting a 3-dimensional vacuum brane along the boundary(see Fig.\ref{fig:Cmetric4D}). Because the string is contained in the removed part of the spacetime, the braneworld solution obtained by this procedure is regular everywhere. 

This example suggests that if there exists a static localised black hole solution in higher-dimensional braneworld model, its analytic extension across the boundary brane would give a solution representing a black hole accelerated by a stringy source. Because a vacuum static solution to the Einstein equations are analytic, such an extension always exists. Further, the extended solution cannot be a static solution that has a compact horizon and regular everywhere outside the horizon, because of the uniqueness theorem for regular static black holes in the adS case\cite{Anderson.M&Chrusciel&Delay2004A,Kodama.H2004,Chamblin.A&Hawking&Reall2000,Kodama.H2002a}. Hence, the solution must have singularity or non-compact horizon. In a $D$-dimensional bulk spacetime case with $(D-1)$-dimensional brane, it is natural to assume that the solution has a spatial $\SO(D-2)$ symmetry. In this case, the singularity should also have the same symmetry. Because the solution is regular in the original braneworld, the singularity should be confined inside a half of the spacetime. Simplest such a singularity is a stringy one along the half of the symmetry axis as in the case of the four-dimensional C-metric. Although it is not the most general, we can find a coordinate system in which the singular region is squashed into a singular string, even in more generic cases. Of course, we cannot construct such solutions exactly. However, in the small mass limit of the black hole, it is expected that we can construct such a solution as a perturbation from the $D$-dimensional Schwarzschild-Tangherlini solution because in four dimensions, the braneworld black hole solution constructed from the C-metric approaches the Schwarzschild solution in the same limit.

On the basis of these observations, in the present paper, we construct solutions representing a higher-dimensional black hole pulled by a stringy source with a small acceleration  utilising the master equation with source for perturbations of the  Schwarzschild-Tangherlini solution developed by the author and his collaborator\cite{Kodama.H&Ishibashi2003,Kodama.H&Ishibashi2004}. Then, we study whether we can construct a localised braneworld black hole solution in the perturbative sense from this solution. Although we cannot give the final answer to the existence and uniqueness of a localised static braneworld black hole with small mass in higher dimensions, we will get some interesting partial results.

The paper is organised as follows. First, in the next section, we briefly review the basic features of the four-dimensional C-metric and its relevance to the braneworld black hole problem and discuss its small acceleration limit. Next, in \S\ref{sec:MasterEq}, we derive a master equation with generic source for static perturbations of the  Schwarzschild-Tangherlini black hole by specialising the general gauge-invariant formulation.  Then, in \S\ref{sec:CmetricSource}, we reconstruct the source term for the perturbative C-metric utilising this master equation to see its structure, and in \S\ref{sec:HDsol} we construct the perturbative accelerated black hole solution by solving the master equation for a stringy source and show that it is uniquely determined by a single function representing the local tension of the stringy source. We also examine the global structure of the solution and show that the stringy singularity is enclosed by a tubular horizon extending to infinity. Finally, in \S\ref{sec:BraneBH}, we study whether there exists a hypersurface satisfying the vacuum brane condition in our accelerated black hole solution. Section \S\ref{sec:summary} is devoted to summary and discussions.

\section{The Four-Dimensional C-metric as a Perturbation}
\label{sec:4DC}

In this section, we consider the small acceleration limit of the C-metric in four dimensions. We can regard the deviation of the metric in this limit from the Schwarzschild metric as a linear perturbation generated by some source in the first order with respect to the acceleration parameter. 

\subsection{C-metric}
\label{subsec:C-metric}

The general C-metric can be expressed as\cite{Plebanski.J&Demianski1976}
\Eqrsub{
&& ds^2=\frac{1}{A^2(x-y)^2}\left[H(y)dt^2-\frac{dy^2}{H(y)}
       +\frac{dx^2}{G(x)}+G(x)d\phi^2\right];
       \label{C-metric:general}\\
&& H(y):=-\nu - K y^2 -2 M A y^3,\\
&& G(x):= 1 - K x^2 -2M A x^3
}
and is a solution to the vacuum Einstein equation with the cosmological constant
\Eq{
\Lambda=-3A^2(\nu+1)=3\lambda. 
}
Note that by an appropriate redefinition of the coordinates $x$ and $y$, we can always set $M\ge0, A\ge0$ and $K=0,\pm1$.

Let us consider the case in which $K=1$ and $MA< 1/(3\sqrt{3})$. In this case, $G(x)$ has three distinct real roots as
\Eq{
G(x)=-2MA(x-x_+)(x-x_-)(x-x_0),
}
and these roots satisfy the inequalities
\Eq{
x_0< -\frac{1}{3MA} < x_- < -1,\quad 0< x_+ < 1.
}
In terms of the parameter $\epsilon$ defined by
\Eq{
M A =\frac{\epsilon}{(1+4\epsilon^2)^{3/2}};\quad 0\le\epsilon <\frac{1}{2\sqrt{2}},
}
these roots are parametrised as
\Eqrsub{
&& x_0=-\frac{\sqrt{1+4\epsilon^2}}{2\epsilon},\\
&& x_+ + x_-=-2\epsilon\sqrt{1+4\epsilon^2},\\
&& x_+ - x_-=2\sqrt{(1+\epsilon^2)(1+4\epsilon^2)}.
}

Let us transform the coordinates $x$ and $y$ to the new coordinates $r$ and $\theta$ defined by
\Eq{
x=x_- + \frac{x_+ - x_-}{2}(1+\cos\theta),\quad
y=- \frac{1}{Ar}.
}
Then, $G(x)$ can be written
\Eq{
G(x)=(1+\epsilon^2)\left(1+2\epsilon g(\theta)\right)\sin^2\theta,
}
where
\Eq{
g(\theta) :=-\epsilon +\sqrt{1+\epsilon^2}\cos\theta.
}
Hence, after rescaling $t$ and $\phi$ as
\Eq{
At \tend t,\quad
\sqrt{\frac{1+\epsilon^2}{1+4\epsilon^2}}
  (1+2\epsilon g(0)) \phi 
 \tend \phi,
}
the C-metric with $K=1$ can be written in terms of the new coordinates as
\Eqr{
& ds^2 =
 & \frac{1}{(1+Ax r)^2}\Big[ -\bar f(r) dt^2+ \frac{dr^2}{\bar f(r)}
\notag\\
&&  +(1+4\epsilon^2)r^2\left\{\frac{d\theta^2}{1+2\epsilon g(\theta)}
    +\frac{1+2\epsilon g(\theta)}{(1+2\epsilon g(0))^2}\sin^2\theta
     d\phi^2\right\}\Big],
\label{C-metric:Sform}
}
where
\Eqr{
&& \bar f(r):= f(r)-\frac{\epsilon^2 r^2}{(1+4\epsilon^2)^3 M^2},\\
&& f(r):= 1-\frac{2M}{r}-\lambda r^2.
}

Clearly, this metric approaches the Schwarzschild(-dS/adS) metric in the limit $\epsilon\tend0$ with $M$ kept constant. Further, for finite $A$, the spacetime is regular in the region with $Axr+1>0$ and $\b f(r)>0$ except on the half of the symmetry axis corresponding to $\theta=\pi$. On this part of the axis, however, the spacetime has a conical singularity represented by the deficit angle 
\Eq{
\Delta\phi= - \frac{4\pi \epsilon (g(0)-g(\pi))}{1+2\epsilon g(0)}
 \approx - 8\pi \epsilon \ (\epsilon\ll1).
}
As is well-known, such a conical singularity is created when there exists a string source with constant line energy density $\mu=-\Delta\phi/(8\pi G) (\approx \epsilon/G)$ and tension $\tau=\mu$. For this reason, the C-metric is regarded as representing a black hole of mass $M$ accelerated by a half-infinite string. In this picture, $\epsilon$ represents the magnitude of acceleration of the black hole.

\subsection{Braneworld black hole}

At $x=0$, the derivative of the metric \eqref{C-metric:general} with respect $x$ is proportional to the metric itself because $G(x)$ has no linear term in $x$. In particular, the extrinsic curvature of the $x=0$ hypersurface can be written
\Eq{
K^\mu_\nu = -\frac{A|y|}{2} g^{\mu\alpha}\frac{d}{dx} g_{\alpha\nu}=A h^\mu_\nu,
}
where $h_{\mu\nu}$ is the induced metric on the hypersurface $x=0$. This is identical to the Israel junction condition for a three-dimensional vacuum brane with positive tension $\kappa^2 T^\mu_\nu=-4Ah^\mu_\nu$ in the 4-dimensional bulk. Hence, we obtain a braneworld black hole solution  if we cut off the $x<0$ part and put a brane with positive tension  at $x=0$, as first pointed out by Emparan, Horowitz and Myers\cite{Emparan.R&Horowitz&Myers2000}. Because the string singularity is contained in the region $x<0$, the corresponding solution is regular. Further, if we choose $\lambda$ so that $\lambda=-A^2$, the 3-dimensional spacetime on the brane become asymptotically flat and has a horizon at $r=2M$. This parameter choice corresponds to $\nu=0$.

\subsection{$\epsilon$-expansion}

If we consider the $\epsilon\tend0$ limit with finite fixed $M$ in this model, the corresponding braneworld black hole solution can be regarded as a perturbation of a Schwarzschild black hole up to the linear order in $\epsilon$, because $\lambda=\Order{\epsilon^2}$. 
From the expansion of the C-metric with respect to $\epsilon$,
\Eqr{
& ds^2= 
 & \left(1-2\frac{\epsilon}{M}r\cos\theta\right)
   \Big[-f(r)dt^2+ \frac{dr^2}{f(r)} \notag\\
&& \quad +r^2\left\{(1-2\epsilon\cos\theta) d\theta^2
         +[1-2(2-\cos\theta)\epsilon]\sin^2\theta d\phi^2\right\}\Big],
}
the explicit expressions for the metric perturbation, $h_{\mu\nu}=\delta g_{\mu\nu}$, is given by
\Eqrsubl{D=4:hmn}{
&& h^t_t=-\frac{2\epsilon}{M}r\cos\theta,\quad
   h^r_r=-\frac{2\epsilon}{M}r\cos\theta,\quad
   h^r_t=0,\\
&& h_{ai}=0,\\
&& h^\theta_\theta=-2\epsilon\left(1+\frac{r}{M}\right)\cos\theta,\ 
   h^\phi_\phi=-2\epsilon
        \left[2+\left(\frac{r}{M}-1\right)\cos\theta\right],\ 
   h_{\theta\phi}=0.
}
In particular, if we decompose the angular part $h_{ij}$ as 
\Eq{
h_{ij}= 2r^2 ( h_L \gamma_{ij} + h_{Tij}),
}
the trace $h_L$ and the traceless part $h_{Tij}$ are given by
\Eqrsub{
&& h_L=-\epsilon -\frac{\epsilon r}{M}\cos\theta,\\
&& \left((h_T)_{\theta\theta},(h_T)_{\theta\phi},(h_T)_{\phi\phi}\right)
   =(1-\cos\theta )\epsilon \times \inpare{1,0,-\sin^2\theta}.
}

After developing a general gauge-invariant formulation for perturbations of the Schwarzschild black hole in arbitrary dimensions, we will show that the source energy-momentum tensor obtained by inserting these expressions into the perturbative Einstein equations is given by $\kappa^2\delta T^a_b = -{8\pi\epsilon}/{r^2}\delta^2(-\Omega) \delta^a_b$, which coincides with the energy-momentum tensor for a half-infinite string with constant line density $\mu=8\pi \epsilon/\kappa^2$ put on the symmetry axis. 

\section{The Master Equation for Scalar Perturbations with Source}
\label{sec:MasterEq}

In the present paper, we generalise the above perturbative analysis of the C metric to higher dimensions to construct a class of perturbative solutions that can be regarded as representing a black hole accelerated by a straight string. For that purpose, in this section, we derive a master equation for such perturbations by specialising the general gauge-invariant formulation for perturbations with source in a higher-dimensional static black hole background developed in Ref. \citen{Kodama.H&Ishibashi2004} to static perturbations in the Schwarzschild-Tangherlini background,
\Eqr{
&& ds^2= -f(r) dt^2 + \frac{dr^2}{f(r)} + r^2d\Omega_n^2,\\
&& f(r)= 1-x;\quad x=\pfrac{r_h}{r}^{n-1},
}
where $d\Omega_n^2=\gamma_{ij} dz^i dz^j$ is the metric of the unit Euclidean $n$-sphere. The dimension of the whole spacetime $D$ is given by $D=n+2$. 

It is clear that perturbations relevant to this problem can be assumed to be invariant under the $\SO(n)$ group representing rotations around the string in $(n+2)$-dimensional spacetime. As shown in Appendix \ref{sec:SOStensors}, such perturbations are of the scalar type if we require that perturbations are regular in one half of the spacetime region outside the horizon. Hence, in the present paper, we only consider the scalar-type perturbation.

\subsection{Perturbation variables}

A scalar perturbation of the metric can be expanded in terms of the scalar harmonics $\SHB$ on the unit sphere $S^n$ as
\Eq{
\delta g_{ab}=f_{ab}\SHB,\quad 
\delta g_{ai}=r f_a \SHB_i,\quad
\delta g_{ij}=2r^2 (H_L \gamma_{ij} \SHB + H_T \SHB_{ij}).
}
(See Appendix \ref{sec:HarmonicTensors} for the basic definitions and properties of tensor harmonics on $S^n$.) 
A natural basis of the gauge-invariant variables for the metric perturbation is given by\cite{Kodama.H&Ishibashi&Seto2000}
\Eqrsubl{GaugeInvVar:scalar:metric}{
&& F=H_L+\frac{1}{n}H_T+\frac{1}{r}D^ar X_a,\\
&& F_{ab}=f_{ab}+D_aX_b+D_bX_a,
}
with 
\Eq{
X_a=\frac{r}{k}\left(f_a+\frac{r}{k}D_a H_T\right).
\label{Xa:def}
}
Here and in the following, the indices $a,b,\cdots$ represent either $t$ or $r$, and  $i,j,\cdots$ correspond to the coordinates of $S^n$. $D_a$ is the covariant derivative with respect to the 2-dimensional metric $g_{ab}dx^a dx^b=-f dt^2+ dr^2/f$. $k$ in the definition for $X_a$ is related to the eigenvalue of the harmonics on $S^n$ as $\hat\triangle_n \SHB=-k^2 \SHB$. To be explicit, $k^2$ takes the discrete values $k^2=l(l+n-1)$ ($l=0,1,\cdots$).

Note that $\SHB_i$ and $\SHB_{ij}$ effectively vanish for $l=0$. The factor $1/k$ and $1/k^2$ in their definitions are introduced for convenience and are not essential. Hence, we have to put $f_a=H_T=0$ for this mode. Similarly, $\SHB_{ij}$ vanishes for $l=1$\cite{Kodama.H2007Aa}. We also have to put $H_T=0$ in this case. These modes with $l=0,1$ are called the exceptional modes. For these modes, $F$ and $F_{ab}$ are not gauge invariant. Here, we give their transformation laws only for the static case relevant to the present paper. For $l=0$, they are given in terms of two functions $T_a(r)$ ($a=t,r$) and a constant $\alpha$ as
\Eq{
\b\delta F=-\frac{1}{r}T^r,\ 
\b\delta F^t_t=2\alpha-f' T_r,\ 
\b\delta F^r_r=-2f T_r'-f'T_r,\ 
\b\delta F^t_r=-(T^t)',
\label{GaugeTrf:l=0:F}}
and for $l=1$, they are given in terms of a single function $L(r)$ as
\Eq{
\b\delta F=-\frac{r}{k}\inpare{fL'+\frac{L}{r}},\ 
\b\delta F^t_t=-\frac{r^2}{k}f'L',\ 
\b\delta F^r_r=-\frac{r}{k}\inpare{2rfL''+(rf'+4f)L'},\ 
\b\delta F^t_r=0.
\label{GaugeTrf:l=1:F}}
Here, the prime denotes the differentiation with respect to $r$.

Next, for a scalar perturbation of the energy-momentum tensor expressed as
\Eq{
\delta T_{ab}= \tau_{ab}\SHB,\quad
\delta T^a_i=r \tau^a \SHB_i,\quad
\delta T^i_j=\delta P \delta^i_j\SHB + \tau_T \SHB^i_j,
}
$\tau_T$ and the following combinations provide a gauge-invariant basis\cite{Kodama.H&Ishibashi&Seto2000}:
\Eqrsubl{GaugeInvVar:scalar:EMtensor}{
&& \Sigma_{ab}=\tau_{ab}-P(D_aX_b+D_bX_a)-X^cD_c P g_{ab},
\label{Sigma_ab:def}\\
&& \Sigma_a=\tau_a +\frac{2k}{r}PX_a,
\label{Sigma_a:def}\\
&& \Sigma_L = \delta {P} +X^aD_a{P},
\label{Sigma_L:def}
}
where $P$ is the background pressure defined by $T^i_j=P \delta^i_j$. Note that for the vacuum background as considered in the present paper, all components $\delta T_{\mu\nu}$ are gauge invariant by themselves. As for the metric perturbation variables, $\Sigma_a$ for $l=0$ and $\tau_T$ for $l=0,1$ are not defined.

\subsection{The Einstein equations}

In order to derive a master equation for scalar perturbations, it is convenient to introduce the four variables $X,Y,Z$ and $S_T$ defined by
\Eqrsubl{XYZ:def}{
&& X:=r^{n-2}(F^t_t-2F),\ 
  Y:=r^{n-2}(F^r_r-2F),\ 
  Z:=r^{n-2} F^r_t,\\
&& S_T:=-r^{n-2}[F^a_a+2(n-2)F].
}
The original metric variables $F^a_b$ and $F$ are expressed in terms of these as
\Eqrsub{
&& r^{n-2}F=-\frac{1}{2n}(X+Y+S_T),\\
&& n r^{n-2}F^t_t= (n-1)X -Y -S_T,\\
&& n r^{n-2}F^r_r= -X +(n-1)Y -S_T,\\
&& r^{n-2}F^r_t=Z.
}
Further, in order to make the final expressions simpler, we rescale the metric perturbation variables other than $\tau_T$ as
\Eq{
S_{ab}:=r^{n-2}\kappa^2\Sigma_{ab},\quad 
S_a:=\frac{r^{n-1}}{k}\kappa^2\Sigma_a,\quad 
S_L:= r^{n-2}\kappa^2 \Sigma_L.
}

In terms of these variables, the Einstein equations for time-independent scalar perturbations can be shown to be equivalent to the set
\Eqrsubl{EinsteinEqs}{
&E_t:& Z'=-2S_t,
\label{PEEqs:static:Et}\\
&E_r:& -Y'+\frac{f'}{2f}(X-Y)=2S_r+r^{n-1}\left(S_T/r^{n-1}\right)',
\label{PEEqs:static:Er}\\
&E^r_t:& \frac{k^2}{r^2}Z=2S^r_t,
\label{PEEqs:static:Ert}\\
&E^r_r:& -r^2 f'X'-\left(rf'+2nf\right) rY'
 \notag\\
&& +(n-1)(n-2)x X+[2m-(n+1)(n-2)x]Y
 \notag\\
&& -(rf'+2nf)rS_T'+[2m+2n(n-1)-(n+1)(n-2)x]S_T
 \notag\\
&&=4r^2S^r_r,
\label{PEEqs:static:Err}\\
&E_T:& S_T=\frac{2r^{n}}{k^2}\kappa^2\tau_T,
\label{PEEqs:static:ET}
}
and the perturbation of the energy-momentum conservation laws
\Eqrsubl{EMconsEqs}{
&& (r^2 f S_r)'-r^2 S_L+\frac{(n-1)m}{2n}S_T=0,
\label{EMcons:r}\\
&& (r^2 S^r_t)'+ k^2 S_t=0,
\label{EMcons:rt}\\
&& (r^2 S^r_r)'+\frac{r^2 f'}{2f}(S^r_r-S^t_t)+k^2 S_r -nr S_L=0.
\label{EMcons:rr}
}
Here, $m$ is related to the eigenvalue $k^2$ of the corresponding as 
\Eq{
m:= k^2 - n = (l-1)(l+n),\quad l=0,1,2,3,\cdots.
}

For uniform treatments, we regard \eqref{PEEqs:static:ET} as the definition of $\tau_T$ for the exceptional modes with $l=0$ or $1$, and \eqref{PEEqs:static:Et} and \eqref{PEEqs:static:Er} as definitions for $S_t$ and $S_r$ for $l=0$, respectively. \eqref{EMcons:r} does not appear for $l=0$,

A spacetime metric is static if its expression is independent of the time variable $t$ and in addition it is invariant under the time inversion $t\tend -t$. Hence, we require that the metric perturbation variables satisfy
\Eq{
Z= r^{n-2}F^r_t=0.
}
Then, the above Einstein equations require that the energy-momentum tensor satisfies
\Eq{
S_t= S^r_t=0.
}
The remaining non-trivial components of the Einstein equations, $E_r$ and $E^r_r$, gives a first-order system of ODEs for $X$ and $Y$ with respect to $r$. It is easy to reduce this set to a second-order ODE for $Y$ with respect to $x=(r_h/r)^{n-1}$:
\Eqr{
&& \frac{d^2 Y}{dx^2}+P \frac{dY}{dx} + Q Y =S_Y;
\label{Yeq:static:general}\\
&& S_Y:= -\frac{2r^2(S^r_r+S_L)}{(n-1)^2x^2(1-x)} 
       -\mdiff{2}{S_T}{x} -P_1\diff{S_T}{x}- Q_1 S_T,
\label{S_Y:general}
}
where
\Eqrsub{
&& P:=-\frac{2[1+(n-2)x]}{(n-1)x(1-x)},\\
&& Q:=\frac{-m+(n-2)x}{(n-1)^2x^2(1-x)},\\
&& P_1:=-\frac{(5n-9)x-2n+6}{2(n-1)x(1-x)},\\
&& Q_1:=-\frac{n(n-2)(n-3)x+2n(n^2-1)+2(2n-1)m}
          {2n(n-1)^2 (1-x)x^2}.
}

For each solution to this equation, $X$ for $l\ge1$ is determined from \eqref{PEEqs:static:Er} as 
\Eq{
X=Y-2f \frac{dY}{dx}+\frac{4f}{(n-1)x}rS_r
 -2f\inpare{\frac{dS_T}{dx}+\frac{S_T}{x}}.
\label{XbyY:general}}
The residual gauge freedom for the exceptional modes with $l=1$ is expressed in terms of $X$, $Y$ and $S_T$ as
\Eqrsubl{GaugeTrf:Y&ST:l=1}{
\b\delta X &=& \frac{r^{n-2}}{k}\insbra{\inpare{-rf'+2f}rL'+2L},\\
\b\delta Y &=& -\frac{r^{n-2}}{k}\insbra{2r^2fL''+\inpare{2f+rf'}
r L'-2L},\\
\b\delta S_T &=&
 \frac{2}{k}\insbra{(r^n f L')' + (n-2)r^{n-2}L}.
}

\eqref{XbyY:general} holds only for $l\ge1$ because \eqref{PEEqs:static:Er} does not appear for $l=0$. In order to determine $X$ for the exceptional mode with $l=0$, we have to use the component $E_L$ of the perturbed Einstein equations corresponding to $G^i_i$ (see Ref. \citen{Kodama.H&Ishibashi2004} for the general form of this equation). This equation reads
\Eqr{
&& \frac{d}{dx}\insbra{x^{(n-2)/(n-1)}\inrbra{X-\frac{(n+1)x-2n}{(n-1)x}(Y+S_T)}}
\notag\\
&&\quad
 =\frac{1}{(n-1)^2 x^{n/(n-1)}}\inrbra{4r^2 S^r_r-2n(n-3)S_T}.
}
From this, we can determine $X$ up to an integration constant. The residual gauge freedom for this mode can be expressed as
\Eqrsubl{GaugeTrf:Y&ST:l=0}{
&& \b\delta X=2\alpha r^{n-2}+r^{n-3}(-rf'+2f)T_r,\\
&& \b\delta Y = -2r^{n-1}\sqrt{f}\inpare{\frac{\sqrt{f}}{r}T_r}',\\
&& \b\delta S_T = -2\alpha r^{n-2}+ 2 \inpare{r^{n-2}f T_r}',
}
%

\section{The Source Term of the Perturbative C-metric}
\label{sec:CmetricSource}

Before considering the general solution of the master equation derived in the previous section, let us calculate the values of the basic gauge-invariant variables $F$ and $F_{ab}$ ($a,b=t,r$) for the metric perturbation \eqref{D=4:hmn} and then, using the Einstein equations, determine the gauge-invariant source variables $S^a_b, S_a$, $S_L$ and $\tau_T$ for the perturbative C-metric. 

\subsection{The $l=0$ mode}
\label{sec:CmetricSource:l=0}

First, the spherically symmetric component of the metric perturbation reads
\Eq{
f^{(0)}{}^a_b=0,\quad  H^{(0)}_L=-\epsilon,
}
which leads to
\Eq{
F=-\epsilon,\quad
F^a_b=0.
}
By inserting this into \eqref{XYZ:def}, \eqref{EinsteinEqs} and \eqref{EMconsEqs}, we obtain
\Eq{
X=Y= 2\epsilon,
}
and
\Eq{
S^a_b= -\frac{2\epsilon}{r^2}\delta^a_b,\quad
S_a=0,\quad
S_L=0,\quad
S_T=0.
}
As mentioned in the previous section, $S_a$ and $S_T$ are not gauge invariant for this mode, although their values do not affect the source energy-momentum tensor $T_{\mu\nu}$. Further, even if we impose the gauge condition $S_a=S_T=0$, there remains the residual gauge freedom given by 
\Eq{
T_r=\frac{1}{f}\inpare{\alpha r+\beta},
}
where $\beta$ is a constant, from  \eqref{GaugeTrf:Y&ST:l=0}. This transforms $X$ and $Y$ to 
\Eqrsub{
&& X \then 2\epsilon +2\alpha + \frac{2-3x}{1-x}\frac{2M\alpha+\beta x}{2M},\\
&& Y\then 2\epsilon +\frac{2\beta}{r} + \frac{x}{1-x}\frac{2M\alpha+\beta x}{2M}
}
If we require the regularity of $Y$ at horizon, $\beta$ is determined in terms of $\alpha$ as $\beta=-2M\alpha$, and we are left with residual gauge freedom parameterised by the single constant $\alpha$. This residual transformation corresponds to a kind of scaling transformations of coordinates $t$ and $r$.

\subsection{The $l=1$ modes}

From Appendix A, the $\SO(2)$-symmetric harmonics with $l=1$ are given by
\Eqrsub{
&& \SHB^{(1)}= \cos(\theta),\\
&& \SHB_\theta^{(1)}=\frac{1}{\sqrt{2}} \sin\theta,\quad
   \SHB_\phi^{(1)}=0,\\
&& \SHB^{(1)}_{ij}=0.
}
Hence, the $l=1$ component of the metric perturbation reads
\Eq{
 f^{(1)}{}^a_b=-2\epsilon\frac{r}{M}\delta^a_b, \quad
   f^{(1)}_a=0,\quad
   H_L^{(1)}=-\epsilon\frac{r}{M},\quad
   H_T^{(1)}=0,
}
and we have
\Eq{
F=-\epsilon \frac{r}{M},\quad
F^a_b = -2\epsilon \frac{r}{M} \delta^a_b.
\label{F:n=2,l=1}}
From \eqref{XYZ:def}, \eqref{EinsteinEqs} and \eqref{EMconsEqs}, these determine $X$ and $Y$ as
\Eq{
X=Y=0,
\label{X,Y:n=2:l=1}
}
and the source terms as
\Eq{
S^a_b= \frac{6\epsilon}{r^2}\delta^a_b,\quad
S_a=0,\quad
S_L=0,\quad
S_T=4\epsilon \frac{r}{M}.
\label{S:n=2:l=1}
}

As for $l=0$, we can change the value of $X$, $Y$ and $S_T$ by a gauge transformation without affecting $T_{\mu\nu}$. From \eqref{GaugeTrf:Y&ST:l=1}, $S_T$ can be put to zero by transformations
\Eq{
L'= -\frac{\epsilon k}{M f}\inpare{1-\frac{C_1}{r^2}},
}
where $C_1$ is an arbitrary constant. We can easily check that $Y$ can be transformed into an expression that is regular at horizon only when we take $C_1$ to be $C_1=r_h^2=4M^2$. For  this choice, $Y$ is transformed to
\Eq{
Y = \epsilon\inpare{-4\ln\frac{r}{2M}+\frac{12M}{r}}+C,
}
where $C$ is an arbitrary gauge parameter. Note that we can go to this gauge preserving the properties $f_{rt}=f_a=0$ from the gauge transformation law
\Eq{
\bar\delta f_{rt}=-f(T_t/f)',\quad
\bar\delta f_t=\frac{k}{r}T_t,\quad
\bar\delta f_r=-rL'+\frac{k}{r}T_r.
}

Although $Y$ can be made regular at horizon, $Y$ grows logarithmically with $r$ at $r=\infty$. This behavior is related to the linear growth of the metric perturbation variables in $r$, \eqref{F:n=2,l=1}. These linear terms cannot be simultaneously eliminated  by a gauge transformation. In fact, in this new gauge, we have
\Eqrsub{
&& F=\epsilon\inpare{-\frac{r}{M}-1+2\ln\frac{r}{2M}}-\frac{C}{2},\\
&& F^t_t=-F^r_r= 2\epsilon\inpare{\frac{r}{M}+1-\frac{12M}{r}}.
}
%

\subsection{$l\ge2$ modes}

The $\SO(2)$-symmetric harmonic functions with $l\ge2$ are given by $\SHB^{(l)}=P_l(\cos\theta)$, which satisfy the normalisation condition
\Eq{
\int_{S^2}d\Omega |\SHB^{(l)}|^2=\frac{4\pi}{2l+1}.
}
The corresponding tensor harmonics $\SHB^{(l)}_{ij}$ satisfies the normalisation condition 
\Eq{
\int_{S^2} d\Omega\, \SHB^{(l)}_{ij} \SHB^{(l)ij}
=\frac{2\pi(l-1)(l+2)}{l(l+1)(2l+1)}.
}

By expanding the $l\ge2$ part of the metric perturbation,
\Eq{
h^a_b=h_{ai}=0,\quad
h^\theta_\theta=-h^\phi_\phi=2\epsilon (1-\cos\theta),\quad
h^\theta_\phi=0,
}
in terms of these tensor harmonics, we obtain
\Eq{
f^{(l)}{}^a_b=0, \quad
f^{(l)}_a=0,\quad
H_L^{(l)}=0,\quad
H_T^{(l)}=\frac{4 \epsilon(2l+1)}{(l-1)(l+2)} (-1)^l.
}
The corresponding gauge-invariant variables are
\Eq{
F=\frac{1}{2}H_T,\quad
F^a_b=0.
}
In terms of $X, Y$ and $S_T$, these are expressed as
\Eq{
X=Y=-H_T,\quad S_T=0,
\label{X,Y,S_T:n=2:l>=2}
}
and the source terms are determined as
\Eq{
S^a_b=-\frac{2\epsilon(2l+1)}{r^2}(-1)^l \delta^a_b,\quad
S_a=S_L=\tau_T=0.
}
%

\subsection{Source distribution}

The energy-momentum tensor $T^\mu_\nu(x)$ of the source can be determined by summing up all of its harmonic components determined so far. Since $S_a, S_L$ and $S_T$ vanish for all modes except $S_T$ for $l=1$, which does not contribute to $T^\mu_\nu$, $T^a_i$ and $T^i_j$ vanish identically. Hence, the only non-trivial components are
\Eq{
\kappa^2 T^a_b(r,\Omega)
 =\delta^a_b\frac{2\epsilon}{r^2}\sum_{l=0}^\infty
     (-1)^{l-1}(2l+1)P_l(\cos\theta).
}
From the formula \eqref{deltabyCl} with $P_l=C^{1/2}_l$, they can be written
\Eq{
\kappa^2\delta T^a_b = -\frac{8\pi\epsilon}{r^2}
   \delta^2(-\Omega) \delta^a_b.
}
This coincides with the energy-momentum tensor for a half-infinite string with constant line density $\mu=8\pi \epsilon/\kappa^2$ put on the south half of the symmetry axis. This result is consistent with the stringy interpretation of the singularity of the C-metric given in \S\ref{subsec:C-metric}.

\section{Solutions for a Stringy Source}
\label{sec:HDsol}

In this section, we construct the general solution to the master equation for a stringy source and study its basic properties.

\subsection{General solution}

When $2lp$ for $p=1/(n-1)$ is not an odd integer, the fundamental solutions to the master equation \eqref{Yeq:static:general} with vanishing source terms are given by 
\Eq{
\frac{x^{1+p(l+1)}}{1-x} F_1(x),\quad
\frac{x^{-p(l-1)}}{1-x} F_2(x),
}
where $F_1(x)$ and $F_2(x)$ are expressed in terms of  the hypergeometric function as
\Eq{
F_1(x)= F(lp,lp+1,2lp+2;x),\quad F_2(x)= F(-lp,-lp-1,-2lp;x).
}
When $2l=(2m+1)(n-1)$ ($m=0,1,2,\cdots$), which occurs only when $n$ is odd, $F_2$ should be replaced by
\Eq{
F_2(x)= F^*(x)- C x^{2m+2}F_1(x)\ln x,
\label{F2:2lp:odd}
}
where
\Eqr{
C&=& \frac{(-lp)_{2m+2}(-lp-1)_{2m+2}}{(2m+1)!(2m+2)!},\\
F^*(x) &=& \sum_{j=0}^{2m+1} (-1)^j \frac{(2m+1-j)!(-lp)_j(-lp-1)_j}{(2m+1)!}x^j
\notag\\
&& -C x^{2m+1}\sum_{j=1}^\infty x^j \frac{(lp+1)_j (lp)_j}{j!(2m+3)_j}
 \notag\\
&&\quad\times
 \sum_{i=1}^j \inpare{-\frac{1}{lp+i}-\frac{1}{lp+i-1}
  +\frac{1}{2m+i+2}+\frac{1}{i}},
}
with
\Eq{
(\alpha)_j:=\frac{\Gamma(\alpha+j)}{\Gamma(\alpha)}.
}
Note that $F_1(x)$ and $F_2(x)$ are bounded in the interval $0\le x\le1$ for any value of $lp\ge0$. 

In terms of there fundamental solutions, the general solution to the master equation, $Y^{(l)}$, for the $l$-th eigenvalue can be expressed as
\Eqr{
Y^{(l)} &=& \frac{2(-1)^l}{(n-1)(1-x)}
  \Big[ -x^{1+p(l+1)} F_1(x)\int^1_x dx_1 x_1^{-2-p(l+1)}s_l(x_1) F_2(x_1)
  \notag\\
 && 
 + x^{-p(l-1)}F_2(x) \int^1_x dx_1 x_1^{-1+p(l-1)} s_l(x_1) F_1(x_1)
 \notag\\
 &&
 +A x^{1+p(l+1)}F_1(x) + B x^{-p(l-1)}F_2(x)  \Big],
\label{GeneralSol:Y}
}
where $A$ and $B$ are constants and
\Eq{
s_l(x):= \frac{(-1)^l (n-1)^2 x^2(1-x)}{2(2l+n-1)}S_Y(x).
}
%

\subsection{Stringy source}

In the present paper, we assume that $T^M_N$ has the structure
\Eq{
T^M_N= t^M_N(r)\delta^n(-\Omega).
}
Here, although $t^M_N$ can contain derivatives in the direction orthogonal to the string, which we denote $\theta$ as in Appendix \ref{sec:SOStensors} in general, we do not consider such a  multipole-type source in the present paper. So, $t^M_N$ is a normal function only of $r$ such that $t^t_r=t^t_i=0$. Then, from the $\SO(n)$ symmetry around the string, we have $t^r_i=0, t^\theta_A=0$ and $t^A_B\propto \delta^A_B$, where we have used the same notation as in Appendix \ref{sec:SOStensors} for the angular coordinates perpendicular to the string. This implies that the tracefree part of $T^i_j$ is proportional to $[n-1,-1,\cdots,-1]$, and the inner product of it with a harmonic tensor $\SHB_{ij}$ is proportional to the value of 
\Eq{
\inpare{n\h D^2_\theta-\h\triangle}\SHB=n \pd_\theta^2 \SHB+ k^2\SHB
}
at $\theta=\pi$. This quantity vanishes because from \eqref{Harmonics:SO(n)-symmetric} we have
\Eq{
-k^2\SHB=\inpare{\frac{d^2}{d\theta^2}+(n-1)\cot\theta\frac{d}{d\theta}}\SHB=n\frac{d^2}{d\theta^2}\SHB
}
at $\theta=\pm\pi$. 

Hence, we can write
\Eq{
T^a_b=t^a_b(r)\delta^n(-\Omega),\quad
T^a_i=0,\quad
T^i_j=t_L(r) \delta^i_j \delta^n(-\Omega).
}
From \eqref{deltabyCl}, the harmonic expansion of these expressions yields 
\Eqrsub{
&& S^{(l)}{}^a_b= S^{(0)}{}^a_b(x) \frac{a_l}{a_0},\quad
   S^{(l)}{}_L= S^{(0)}{}_L(x) \frac{a_l}{a_0}\ (l\ge0),\\
&& S^{(l)}{}_t=S^{(l)}{}_r=0\ (l\ge1),\quad
  S^{(l)}{}_T= 0 \ (l\ge2),
}
where $S^{(0)}{}^a_b$ and $S^{(0)}{}_L$ are constant multiples of $r^{n-2}t^a_b$ and $r^{n-2}t_L$, respectively. Inserting these into \eqref{EMcons:r}, we obtain $S_L^{(l)}=0$. Then, \eqref{EMcons:rr} reads
\Eq{
\inpare{r^2S^{(0)}{}^r_r}' + \frac{r^2 f'}{2f} \inpare{S^{(0)}{}^r_r-S^{(0)}{}^t_t}=0.
}
This implies that all source terms are completely  determined by $S^{(0)}{}^r_r$ or equivalently by $s_0(x)$. Further, from the $l$ dependence of $S^{(l)}{}^r_r$ and from \eqref{al}, we find that $s_l$ defined above is independent of $l$,
\Eq{
s_l=s,
}
and the function $t^r_r$ can be written in terms of $s$ as
\Eq{
t^r_r= -\frac{\mu}{r^n};\quad
\kappa^2 \mu(r)=\frac{8\pi \Gamma(n-1)\Omega_{n-1}}{2^n\Gamma\inpare{\frac{n-1}{2}}^2}s(x).
}
Thus, roughly speaking, $s(x)$ characterises the local tension of the stringy source.
The other components of $t^a_b$ are determined as
\Eq{
t^t_t=-\frac{2\sqrt{f}}{r^n f'} \inpare{\mu \sqrt{f}}',\quad
t^t_r=0.
}
%

\subsection{Regularity and asymptotic condition}

\subsubsection{Generic modes}

As shown in \S\ref{sec:CmetricSource}, we can always find a gauge in which the metric perturbation is regular at horizon for the perturbative C-metric in four dimensions. So, we also require the regularity of perturbations at horizon $x=1$ in higher dimensions. Then, $A$ and $B$ in \eqref{GeneralSol:Y} should be related as
\Eq{
A F_1(1)+ B F_2(1)=0.
\label{regularity:horizon}
}

Next, since $F^a_b$ is of the order of $Y^{(l)}/r^{n-2}$ at $r\sim\infty$ along a generic angular direction, we require that $Y^{(l)}$ is bounded at $r\tend \infty$ for $l\ge2$, so that the induced metric on a brane transversal to the black hole exhibits the standard asymptotic behavior $\Order{1/r^{n-2}}$ of a vacuum solution in $(n+1)$-dimensional spacetime. Then, together with \eqref{regularity:horizon}, $A$ and $B$ are uniquely determined as
\Eq{
A=-\frac{F_2(1)}{F_1(1)}B,\quad
B=-\int_0^1 dx x^{-1+p(l-1)} s(x)F_1(x) 
\label{B:value}
}
for $l\ge2$. 

For these values of $A$ and $B$, the values of the harmonic expansion coefficients $X^{(l)}(x)$ and $Y^{(l)}(x)$ with $l\ge2$ at horizon and at infinity are determined as follows. First, from the calculations given in Appendix \ref{sec:XY:horizon:l>=2}, the values at infinity are given  by
\Eqrsubl{XY:infinity:l>=2}{
X^{(l)}(0)&=& \frac{2(-1)^l(2l+n-1)}{(l+n-2)(l+1)}
 \insbra{2s'(0)- \frac{l^2+(n-1)l+2-n}{(n+l)(l-1)}s(0)},\\
Y^{(l)}(0)&=&-2 (-1)^l \frac{2l+n-1}{(l-1)(n+l)} s(0).
}
Note that near $x=0$, $X^{(l)}=X^{(l)}(0)+\Order{x}$ and $Y^{(l)}=Y^{(l)}(0)+\Order{x}$.

Next, in terms of $\h Y_1(x)$ and $\h Y_2(x)$ defined by
\Eq{
\h Y_1=x^{1+p(l+1)}F_1(x),\quad
\h Y_2=x^{-p(l-1)}F_2(x),
}
the value of $Y^{(l)}$ at horizon can be written as
\Eqr{
Y^{(l)}(1) &=& \lim_{x\tend 1-0}\frac{2(-1)^{l+1}}{n-1}
 \frac{d}{dx}\insbra{ A \h Y_1(x)+B \h Y_2(x)}
   \notag\\
  &=& \frac{2(-1)^l B}{(n-1)F_1(1)} W(\h Y_1,\h Y_2)(1).
}
Hence, from
\Eq{
W(\h Y_1,\h Y_2)=\frac{2l+n-1}{n-1}x^{2p},
}
and
\Eq{
X^{(l)}=Y^{(l)}-2(1-x)\frac{dY^{(l)}}{dx},
}
we obtain
\Eq{
X^{(l)}(1)=Y^{(1)}(1)
 = \frac{2(-1)^{l}(2l+n-1)B}{(n-1)^2 F_1(1)},
}
where $B$ is given by \eqref{B:value}.

In contrast to these generic modes, the exceptional modes corresponding to $l=0$ and $l=1$ need special treatments. So, we discuss them separately.

\subsubsection{The $l=0$ mode}

Under the gauge condition $S_T=0$, the general solution for the $l=0$ mode is given by
\Eqrsub{
X^{(0)}&=& \frac{(n+1)x-2n}{(n-1)x} Y^{(0)}
      +\frac{4}{n-1}\frac{S^{(0)}_1(x)}{x}+\frac{C}{x^{1-p}},\\
Y^{(0)}&=& \frac{2}{(n-1)(1-x)}\insbra{-S^{(0)}_2(x)+S^{(0)}_1(x)+Ax^{1+p}+Bx^p},
}
where
\Eq{
S_k^{(0)}(x):=x^{p+k-1} \int_x^1 dy\frac{s(y)}{y^{p+k}}
\quad (k=1,2).
}
This solution becomes regular at the horizon when
\Eq{
A+B=0.
}
Under this condition, the values of $X$ and $Y$ at horizon are given by
\Eq{
X^{(0)}(1)=-\frac{2B}{n-1}+C,\quad
Y^{(0)}(1)= \frac{2B}{n-1}.
}
%

By expanding $S_1(x)$ and $S_2(x)$ around $x=0$ with the help of partial integrations, we obtain for $n>2$
\Eqrsubl{X,Y:l=0:expansion}{
Y^{(0)} &=& \frac{2(n-1)}{n} s(0) 
      +2\inrbra{-s(1)+I_1 +\frac{B}{n-1}}x^p
      +\Order{x},\\
X^{(0)} &=&  \frac{\h C}{x^{1-p}}
  -\frac{2(n-1)}{n}s(0)+\frac{4(n-1)}{n-2}s'(0)
   \notag\\
    && +2\inrbra{\frac{n-3}{n-1}s(1)-2s'(1)-I_1 
       +2 I_2 -\frac{(n+1)B}{(n-1)^2} } x^p 
    \notag\\
    &&+\Order{x},
}
where
\Eqr{
&& \h C:=C-\frac{4n}{(n-1)^2}B + \frac{4}{n-1}\inrbra{s(1)-I_1},\\
&& I_n:=\int_0^1 dx\,x^{-p}\pd_x^n s(x).
}
From this, we find that $X^{(0)}$ and $Y^{(0)}$ are finite at infinity if we choose $C$ such that $\h C=0$.

Hence, we are left with one parameter family of solutions even if we impose the regularity condition at horizon and the asymptotic condition at infinity. This parameter can be regarded as representing the freedom of the mass variation for the following reason.

First, note that in general, perturbations with $l=0$ includes the simple mass perturbation of the Schwarzschild metric,
\Eq{
\delta F=0,\quad
\delta F_{tt}=\frac{2\delta M}{r^{n-1}},\quad
\delta F_{rr}=\frac{1}{f^2} \frac{2\delta M}{r^{n-1}},
}
which changes $X$ and $Y$ as
\Eq{
\delta X=-\frac{2\delta M}{rf},\quad
\delta Y=\frac{2\delta M}{rf}.
}
These transformations are singular at horizon.

In the meantime, $X$, $Y$ and $S_T$ for the $l=0$ mode are not gauge invariant, as mentioned in \S\ref{sec:MasterEq}. From \eqref{GaugeTrf:Y&ST:l=0}, even under the gauge condition $S_T=0$, there remains the residual gauge freedom with $T_r= \frac{C_g}{r^{n-2}f}$, which transforms $X^{(0)}$ and $Y^{(0)}$ as
\Eqrsub{
&& \b\delta X^{(0)}= \frac{2-(n+1)x}{r(1-x)}C_g,\\
&& \b \delta Y^{(0)}=\frac{2-x}{r(1-x)}(n-1)C_g.
}
These transformations are also singular at horizon. However, we can take an appropriate linear combinations of these and the above mass perturbation to construct the regular transformation
\Eq{
\delta  X^{(0)}= C_g \frac{n+1}{r},\quad
\delta Y^{(0)}=C_g \frac{n-1}{r}.
}
This transformation can still be regarded as a mass perturbation. We can change the value of $B$ in the general solution for $l=0$ preserving the regularity condition $A+B=0$. 

One naive method to remove this degree of freedom is to require that the metric perturbation variables $F$ and $F^a_b$ do not contains a term proportional to the static potential, $x=(r_h/r)^{n-1}$. This condition is equivalent to require that  $X^{(0)}$ and $Y^{(0)}$ do not contain a term of order $x^p\propto 1/r$ in the asymptotic expansions. However, from \eqref{X,Y:l=0:expansion}, we find that this condition is fulfilled only when the tension $s(x)$ satisfies the additional condition
\Eq{
-s(1)+s'(1)+\frac{n}{n-1}I_I-I_2=0.
}
Note that terms proportional to $x^p\propto 1/r$ appear in $X$ and $Y$ only for the $l=0$ mode. If this condition is not satisfied, there is no natural way to fix the total mass of the system. This subtlety does not affect the existence argument on the braneworld black hole in the next section because the $l=0$ mode does not affect the extrinsic curvature of a brane.

Next, we discuss the $n=2$ case. In this case, the asymptotic behaviour of $X$ and $Y$ is different for the higher-dimensional cases because $p=1$ and terms proportional to $\log(x)$ appear: 
\Eqrsub{
X^{(0)} &=& C-s(0)+4s'(0)+4s(1)+4\h I_2
   -8B \notag\\
   && +2\inpare{s'(0)-2s''(0)}x\ln x +\Order{x^2\ln x}
   \notag\\
   && +\inpare{-2s(1)-4s'(1)-s(0)-4s'(0)+2s''(0)
   -2 \h I_2+ 6B}x ,\\
Y^{(0)} &=& s(0)-2s'(0)x\ln x + \inpare{s(0)-2s(1)-2\h I_2 +2B}x + \Order{x^2\ln x}.
}
where
\Eq{
\h I_2:=\int_0^1 dx s''(x)\ln x.
}
Thus, $X^{(0)}$ and $Y^{(0)}$, hence $F^a_b{}^{(0)}$ and $F^{(0)}$  are bounded at infinity irrespective of the values of $B$ and $C$. Therefore, we cannot eliminate the freedom in $C$ unlike for $n>2$. However, this parameter has no physical meaning because its value changes as $C\tend C+2\alpha$ by the gauge transformaion as we saw in \S\ref{sec:CmetricSource:l=0}. The remaining parameter corresponds to the mass freedom as in the case of $n>2$. 

Thus we can understand the physical meaning of the parameters of the solution, but there is another new feature for $n=2$. It is the appearance of terms proportional to $x\ln x$. Since the coefficients of these terms depend only on $s(x)$, the boundedness of the metric perturbations requires the additional constraints 
\Eq{
s'(0)=s''(0)=0.
\label{n=2:s(x):constraint1}
}
In the case of the 4D C-metric, this condition is satisfied because $s$ is constant. If we further require that the terms in proportion to $x$, i.e., to $1/r$, can be removed by the gauge transformation explained above, the following additional condition should be satisfied:
\Eq{
\int_0^1 dx (\ln x-x)s''(x)=0.
\label{n=2:s(x):constraint2}
}
%

\subsubsection{The $l=1$ modes}

For $l=1$, the general solution for $Y^{(1)}$ reads
\Eq{
Y^{(1)}= -\frac{2}{(n-1)(1-x)}\insbra{-S^{(1)}_2(x)F_1(x)
 + S^{(1)}_1(x)F_2(x)+Ax^{1+2p}F_1(x) + BF_2(x)},
}
where
\Eqrsub{
S^{(1)}_1(x) &=& \int_x^1 \frac{dy}{y} s(y) F_1(y),\\
S^{(1)}_2(x) &=& x^{1+2p}\int_x^1 dy y^{-2-2p} s(y)F_2(y).
}
The behaviour of $X$ and $Y$  near the horizon for $l=1$ is the same as that for $l\ge2$. In particular, the regularity at horizon is given by  $AF_1(1)+BF_2(1)=0$, and the values of $X$ and $Y$ at horizon are given by
\Eq{
X^{(1)}(1)=Y^{(1)}(1)=-\frac{2(n+1)B}{(n-1)^2F_1(1)}.
}

In contrast, the behavior of perturbations at infinity is quite different because $Y^{(1)}$ and $X^{(1)}$ increase in proportion to $\log r$ and $r^{n-1}$, respectively. This divergence produces terms growing linearly in $r$ in the original metric perturbation variable $F^a_b$. Such behaviour represents the direct effect of acceleration and is expected from the analysis of the 4D C-metric. In this 4D case, these growing term came from $(1+A xr)^{-2}$ in \eqref{C-metric:Sform}. The perturbative treatment of this term is valid only in the region where $|\epsilon (r/M)\cos\theta|\ll1$. Hence, even if there appears the $\log r$ term in $Y^{(1)}$, it does not implies the divergence of the perturbation in the region where the perturbative treatment is valid. 
 
Anyway, we cannot determine $A$ and $B$ for $l=1$ by the boundary condition at infinity unlike for the other modes. However, this feature does not have any physical importance because they are just gauge freedom. In fact, under the gauge condition $S_T=0$, the residual gauge freedom can be written as
\Eq{
\b\delta Y^{(1)}=-\frac{r^{n-2}}{k}\insbra{2r^2f L''+(2f+rf')fL'-2L},
}
where $L(r)$ is an arbitrary solution to 
\Eq{
(r^n fL')'+(n-2)r^{n-2}L=0.
}
From the gauge invariance of the theory and the quantity $S^r_r$, $\b\delta Y^{(1)}$ satisfies the homogeneous ODE for $Y$ associated with \eqref{S_Y:general}. This implies that two constants $A$ and $B$ in \eqref{GeneralSol:Y} can be changed to any values by gauge transformations.

\subsection{Behavior of the metric perturbation}

Now, let us study the behaviour of the metric perturbation variables by summing up the modes. Since we have already studied the asymptotic behaviour of the exceptional modes, the main task is to estimate the sum of the generic modes
\Eq{
\b X:= \sum_{l=2}^\infty 
         X^{(l)}(x) C_l^{(n-1)/2}(\cos\theta),\quad
\b Y:= \sum_{l=2}^\infty 
         Y^{(l)}(x) C_l^{(n-1)/2}(\cos\theta).
}
%


As shown in Appendix \ref{sec:estimation:modesum}, the values of $\b X$ and $\b Y$ at infinity can be written
\Eqrsub{
\b Y(0,\theta) &=& 2(n-1) s(0)\left[
 -\inrbra{\psi(n+1)-\psi(1)-\frac{n}{n+1}}\cos\theta 
   -\frac{1-\cos\theta}{n}
   \right.\notag\\
&& \left.
  +\frac{1}{\cos^{n/2-1}(\theta/2)}
  \partial_\epsilon F\left(\frac{n}{2}+1-\epsilon,
     -\frac{n}{2}+\epsilon,\frac{n}{2};\sin^2\frac{\theta}{2}\right)
     \right],
  \label{Y:infinity}\\
\b X(0,\theta) &=& \b Y(0,\theta) 
  +\inpare{\frac{s'(0)}{n-2} -s(0)}\left[
     n(n-3) -2(n+1)(n-2)\sin^2\frac{\theta}{2}
      \right. \notag\\
 &&\left.
   +\frac{2}{\cos^{n-2}(\theta/2)}
  F\left(\frac{n}{2}-1,-\frac{n}{2}+2,\frac{n}{2};\sin^2\frac{\theta}{2}\right) \right].
  \label{X:infinity}
}

\begin{figure}[b]
\centerline{
\includegraphics*[height=4cm]{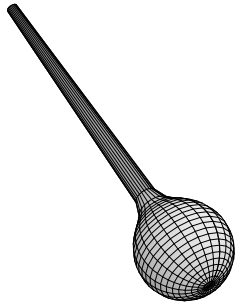}
}
\caption{Tubular horizon of an accelerated black hole}
\label{fig:Blacktail}
\end{figure}

From these, we find that the values of  $\b X$ and $\b Y$ at infinity along the regular part of the symmetry axis are given by
\Eqrsub{
\b Y(0,\theta=0) &=& 2(n-1)s(0)\inpare{\frac{n}{n+1}
 -\psi(n+1)+\psi(1) },\\
\b X(0,\theta=0)&=& (n-1)s(0)
  \inpare{-n+2+\frac{2n}{n+1}-2\psi(n+1)+2\psi(1)}
  \notag\\
&&  +(n-1)s'(0).
}
Further, from
\Eqrsub{
F\inpare{\textstyle \frac{n}{2}-1,-\frac{n}{2}+2,\frac{n}{2};1}
   &=& \frac{\Gamma(n/2)\Gamma(n/2-1)}{\Gamma(n-2)}
     =\frac{\sqrt{\pi}}{2^{n-3}},\\
\pd_\epsilon F\inpare{\textstyle \frac{n}{2}+1-\epsilon,
   -\frac{n}{2}+\epsilon,\frac{n}{2};1} 
   &=&-\frac{\Gamma(n/2)\Gamma(n/2-1)}{\Gamma(n)}
   \notag\\
   &=& -\frac{\sqrt{\pi}}{2^{n-3}(n-1)(n-2)},
}
we have 
\Eqrsub{
\frac{\b Y(x,\theta)}{r^{n-2}}
 &\approx& -\frac{\sqrt{\pi}}{2^{n-4}(n-2)}
           \frac{s(0)}{\rho^{n-2}},\\
\frac{\b X(x,\theta)}{r^{n-2}}
 &\approx& \frac{\sqrt{\pi}}{2^{n-4}(n-2)}
           \frac{s'(0)-(n-1)s(0)}{\rho^{n-2}}
}
near the stringy source, where
\Eq{
\rho:=r \cos\frac{\theta}{2}=r\sin\frac{\pi-\theta}{2}.
}

Note that $\rho$ represents the distance from the stringy source near the source. Further, it is naively expected that the horizon forms where the correction to $g_{tt}$ becomes of order unity, i.e. $F^t_t=\Order{1}$. This condition is equivalent to the condition $s(0)/\rho^{n-2}=\Order{1}$. Hence, the above behavior of the metric perturbation suggests that the horizon is approximately represented as $\rho\approx \const$. That is, the horizon takes a tubular shape that encloses the singular stringy source and extends to infinity, as illustrated in Fig. \ref{fig:Blacktail}.

\section{Application to the Braneworld Black Hole Problem}
\label{sec:BraneBH}

In this section, we study whether we can construct a perturbative braneworld black hole solution from the perturbative accelerated black hole solution obtained in the previous section. The main point is to see whether there exists a hypersurface satisfying the vacuum brane condition
\Eq{
K^\mu_\nu=-\sigma\delta^\mu_\nu
\label{VacuumBraneCondition:general}.
}
%

\subsection{Brane embedding}

For the $\SO(n+1)$-symmetric background
\Eq{
ds^2=-f(r)dt^2+\frac{dr^2}{f(r)}
       +r^2(d\theta^2+\sin^2\theta d\Omega_{n-1}^2),
\label{SSBG}
}
the vacuum brane condition \eqref{VacuumBraneCondition:general} is satisfied only by the equatorial hyperplane\cite{Kodama.H2002a}, for which the tension is given by $\sigma=0$. Therefore, we can assume that a vacuum brane in a perturbed spacetime is located near the plane $\theta=\pi/2$ if it exists:
\Eq{
\theta=\pi/2+\chi(r).
\label{BraneConf}
}

The extrinsic curvature of such a brane is given by\cite{Kodama.H2002a}
\Eq{
\frac{1}{r}K_{\mu\nu}=
  \partial_\mu\partial_\nu \chi +\Gamma^\theta_{\mu\nu}
  -\chi' \Gamma^r_{\mu\nu}+2\Gamma^\theta_{\theta(\mu}\partial_{\nu)}\chi,
}
where $\Gamma^\theta_{\mu\nu}$ should be evaluated for the perturbed metric at the perturbed brane location. 
For static perturbations, the nonvanishing components of this extrinsic curvature can be expressed  in terms of the gauge-invariant variable for the brane location
\Eq{
\hat\chi(r)=\chi(r)-\sum_{k\neq0} \frac{1}{k} H_T\SHB_\theta,
}
as
\Eqrsub{
K^t_t &=& \frac{rf'}{2}\h\chi ' -\frac{1}{2r}\pd_\theta \h h^t_t,\\
K^r_r &=& \frac{f}{r}\inpare{r^2\h\chi'}'+ \frac{rf'}{2}\h\chi'
 -\frac{1}{2r}\pd_\theta \h h^r_r,\\
\frac{K^C_C}{n-1} &=& f\h \chi' + \frac{\h \chi}{r}
 -\frac{1}{r}\pd_\theta \h h_L,
}
where
\Eq{
\h h^t_t=\sum F^t_t \SHB,\quad
\h h^r_r=\sum F^r_r\SHB,\quad
\h h_L=\sum F\SHB ,
}
and the harmonic functions should be evaluated at $\theta=\pi/2$.

Hence, the brane configuration is determined by the set of equations
\Eqrsub{
&& \frac{rf'}{2}\h \chi' -\frac{1}{2r}\partial_\theta \h h^t_t =-\sigma,
\label{BraneEq:1}\\
&& f\h \chi' +\frac{\h \chi}{r}-\frac{1}{r}\pd_\theta\h  h_L =-\sigma,
\label{BraneEq:2}\\
&& rf\h \chi'' +\left(\frac{rf'}{2}+2f\right)\h \chi'
   -\frac{1}{2r}\partial_\theta \h h^r_r=-\sigma.
\label{BraneEq:3}
}
From these equations, we obtain the expression for $\h \chi$ in terms of the metric perturbations,
\Eq{
\h \chi=-\frac{f}{rf'}\partial_\theta \h h^t_t
         +\partial_\theta \h h_L -\left(r-\frac{2f}{f'}\right)\sigma,
\label{Brane:solution}
}
and two constraint equations on the value of perturbation variables on the equatorial hyperplane $\theta=\pi/2$:
\Eqrsub{
&& (\partial_\theta \h h^t_t)'
   +\left(\frac{f'}{2f}-\frac{f''}{f'}\right)\partial_\theta \h h^t_t
   -\frac{f'}{2f}\partial_\theta \h h^r_r 
 =\left(2-\frac{2rf''}{f'}\right)\sigma,
\label{Brane:Constraint1}\\
&& (\partial_\theta \h h_L)' 
   +\left(\frac{f-1}{r^2f'}-\frac{1}{2r}\right)\partial_\theta \h h^t_t
   -\frac{1}{2r}\partial_\theta \h h^r_r
    =\left(\frac{2(f-1)}{rf'}-1\right)\sigma.
\label{Brane:Constraint2}
}

In terms of $Y$, $S_T$, $S_r$ and $S^r_r$, these equation can be expressed as
\Eqrsub{
&& \sum_l \insbra{m(Y+S_T)+n(2-x)rS_r - 2r^2 S^r_r}\pd_\theta C_l^\nu(\pi/2) =2n(n+1)M\sigma,\\
&& \sum_l \insbra{m(Y+S_T)+2nrf S_r - 2r^2 S^r_r}\pd_\theta C_l^\nu(\pi/2) =2n(n+1)M\sigma.
}
Hence, when $S_r=0$, these reduce to the single constraint equation
\Eq{
\sum_l \insbra{m(Y+S_T) - 2r^2 S^r_r}\pd_\theta C_l^\nu(\pi/2) =2n(n+1)M\sigma.
\label{Brane:Constraint}
}
In particular, for the stringy source considered in the present paper, from 
\Eq{
S_T^{(l)}=0,\quad
\sum S^r_r{}^{(l)} C^\nu_l(\cos(\theta))\propto \delta^n(-\Omega),
}
this constraint equation reads
\Eq{
\sum_{l=2}^\infty (l-1)(l+n)Y^{(l)}\pd_\theta C_l^\nu(\pi/2) =2n(n+1)M\sigma.
}
%

\subsection{Source with a constant tension}

Since the source for the C-metric has a constant tension, $r^2 S^r_r=\const$, let us examine whether the brane constraint \eqref{Brane:Constraint} is satisfied for such a source in higher dimensions as well. 

First, we confirm that the constraint is satisfied for the perturbative C-metric. For this metric, from \eqref{X,Y,S_T:n=2:l>=2}, $S_T=0$ and $Y$ is related to the string tension as $mY=2r^2 S^r_r$ for $l\ge2$. Hence, all terms with $l\ge2$ vanish in \eqref{Brane:Constraint}, and from \eqref{X,Y:n=2:l=1} and \eqref{S:n=2:l=1}, the remaining $l=1$ part of \eqref{Brane:Constraint} gives the relation
\Eq{
\sigma=\frac{\epsilon}{M}.
}
With this relation, it is easy to see directly that the expression for the metric perturbation
\Eqrsub{
&& \h h^t_t=\h h^r_r= -2\sigma r \cos\theta,\\
&& \h h_L=-\frac{8\epsilon}{3}\cos\theta\inpare{1+\frac{3}{4}\ln\frac{1+\cos\theta}{2}}-\sigma r\cos\theta
}
satisfies \eqref{Brane:Constraint1} and \eqref{Brane:Constraint2}. In this case, $\h\chi$ is given by
\Eq{
\h\chi=\inpare{\frac{8}{3}-2\ln 2} \epsilon.
}

Now, let us consider the higher-dimensional cases with $n\ge3$. In these cases, for the solution corresponding to the stringy source satisfying 
\Eq{
S_T=0,\quad
r^2 S^r_r=-(-1)^l (2l+n-1)s,
}
the constraint \eqref{Brane:Constraint} reads
\Eq{
\sum_l \insbra{m Y^{(l)}(x) + 2(-1)^l(2l+n-1)s}\pd_\theta C_l^\nu(\pi/2) =2n(n+1)M\sigma.
}
For $x=0$, utilising \eqref{XY:infinity:l>=2} for $l\ge2$, this equation yields 
\Eq{
\sigma=-\frac{s}{nM}.
}
Under this condition, the above constraint can be written 
\Eq{
\sum_{q=1}^\infty (-1)^q \frac{\Gamma(q+\nu+1)}{q!}
 \insbra{ q(2q+1+n) Y^{(2q+1)}(x) -(4q+n+1)s } =0.
}
In particular, from
\Eq{
Y^{(l)}{}'(0)
 =\frac{1}{2}\inpare{Y^{(l)}(0)-X^{(l)}(0)} 
 = -(-1)^l\frac{2(n-2)(2l+n-1)}{(l+1)(l-1)(l+n)(l+n-2)}s,
}
the first derivative of this constraint equation at $x=0$ reads
\Eqr{
0&=&\sum_{q=1}^\infty (-1)^q \frac{\Gamma(q+\nu+1)}{(q-1)!}
  (2q+1+n) Y^{(2q+1)}{}'(0) 
  \notag\\
 &=& (n-2)\sum_{q=1}^\infty (-1)^q\frac{(4q+n+1)\Gamma(q+\nu)}{8(q+1)!}s
  \notag\\
 &=& \frac{(n-2)(n-3)}{16}s
  \sum_{j=0}^\infty
   \frac{n+7+8j}{(2j+3)!}\Gamma(2j+1+\nu).
}
This equation holds only for  $n=2$ or $n=3$. The latter case is also excluded because from \eqref{Y''(0):n=3}, the second derivative of the constraint equations at $x=0$ does not vanish:
\Eq{
\sum_{q=1}^\infty (-1)^q \frac{\Gamma(q+\nu+2)}{(q-1)!} 
Y^{(2q+1)}{}''(0)
 = \frac{3}{2}\sum_{q=1}^\infty \frac{(-1)^{q+1}}{q(q+2)}
   =\frac{3}{8}
}
Thus, we find that in five and higher spacetime dimensions, the accelerated black hole solution with a constant string tension cannot be utilised to construct a localised brane black hole solution, unlike the four-dimensional case.

\subsection{Equation for the tension}

The result in the previous subsection implies that if a localised braneworld black hole can be constructed from an accelerated black hole solution with a stringy source, the tension of that string has to be nonuniform when the bulk dimension is greater than four. 

In order to see whether this generalisation helps or not, we rewrite the constraint equation \eqref{Brane:Constraint} in a form of an integral equation for $s(x)$. First, with the help of the master equation, this constraint equation can be written
\Eq{
\LL_1(x) \left. \pd_\theta \b Y\right|_{\pi/2}
=\frac{2(n+1)(nM\sigma-s(x))}{(n-1)^2x^2(1-x)},
}
where
\Eqr{
\LL_1(x) &:=& \frac{d^2}{dx^2}+P(x)\frac{d}{dx} + Q(m=0)
 \notag\\
 &=& \frac{d^2}{dx^2}
      -\frac{2\inrbra{1+(n-2)x}}{(n-1)x(1-x)}\frac{d}{dx}
         +\frac{n-2} {(n-1)^2x(1-x)}.
}
Because $\b Y$ is expressed in terms of the tension $s(x)$, this equation yields the following integral equation for $s(x)$:
\Eq{
\int_0^1 dy K_n(x,y) s(y)
=\frac{2(n+1)\inpare{n M\sigma-s(x)}}{(n-1)^2x^2(1-x)},
\label{IPfors}
}
where 
\Eqrsub{
K_n(x,y) &=& \LL_1(x) G_n(x,y),\\
G_n(x,y) &=& \frac{2}{(n-1)(1-x)}\sum_{l=2}^\infty
\left[\theta(y-x)\frac{1}{y}\pfrac{x}{y}^{1+p(l+1)} F_1(x)F_2(y)
\right.\notag\\
&&\quad
 +\theta(x-y)\frac{1}{y}\pfrac{y}{x}^{p(l-1)}F_2(x)F_1(y)
 \notag\\
&& \left.\quad
 -\frac{F_2(1)}{F_1(1)} \pfrac{x}{y}^{1+p}(xy)^{pl}
   F_1(x)F_1(y)\right] C^\nu_l{}'(0)
\notag\\
}

\begin{figure}[h]
\includegraphics*[height=4cm]{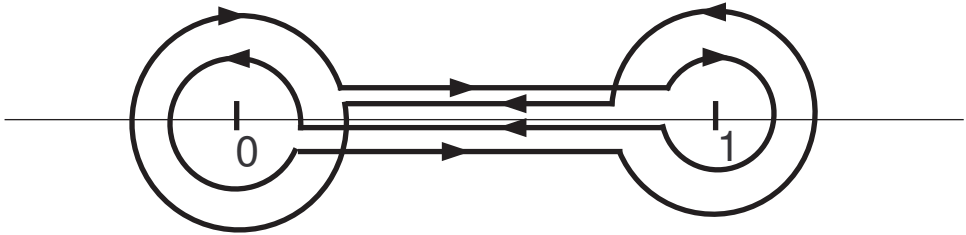}
\label{fig:Contour:C}
\caption{Contour $C$ for $F_2$}
\end{figure}

\begin{figure}[b]
\centerline{
\includegraphics*[width=10cm]{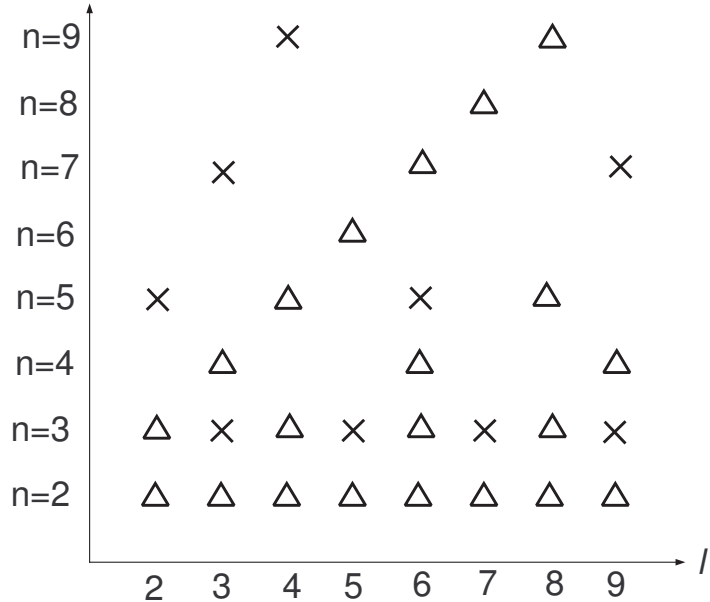}
}
\caption{Special values for $(n,l)$}
\label{fig:DegeneratePoints}
\end{figure}

Now, we rewrite this kernel function utilising the integral expressions for $F_1(x)$ and $F_2(x)$,   
\Eqrsub{
F_1(x) &=& F(lp,lp+1,2lp+2;x) \notag\\
   &=& \frac{\Gamma(2lp+2)}{\Gamma(lp+1)^2}
     \int_0^1 \pfrac{t(1-t)}{1-tx}^{lp} dt,\\
F_2(x) &=& F(-lp,-lp-1,-2lp;x) \notag\\
  &=& \frac{1}{(1-e^{-2\pi lp i})^2}
   \frac{\Gamma(-2lp)}{\Gamma(-lp)^2}
   \int_C \pfrac{1-sx}{s(1-s)}^{lp+1} ds.
}
Here, $C$ in the expression $F_2(x)$ is the contour shown in Fig. \ref{fig:Contour:C} in the complex $s$ plane, and when $lp=k$ is an integer, it is understood that we first put $lp=k+\epsilon$ with $\epsilon\neq0$ and take the limit $\epsilon\tend0$. Note that we cannot use this expression when $2lp$ is an odd integer, in which case we have to use \eqref{F2:2lp:odd} for $F_2(x)$.  These exceptional cases are shown in Fig. \ref{fig:DegeneratePoints} for $2\le n\le 9$ and $2\le l\le 9$. In this figure, a triangle implies that $lp$ is an integer, and a cross implies that $2lp$ is an odd integer. Hence, taking account of the fact that we use these expressions only for odd $l$, the expressions from this point are valid for $n\neq 4j+3$ ($j=0,1,\cdots$).

For such a value of $n$, inserting these integral expressions into the definition of $G_n(x,y)$, we obtain
\Eqr{
&& G_n(x,y) =  \frac{2}{(n-1)(1-x)}\pfrac{x}{y}^{1+p}
\sum_{l=2}^\infty
\left[\frac{(2lp+1)e^{2\pi lp i}}{4\pi  \sin(2lp\pi)}
\right.\notag\\
&&\quad\times
\int_0^1 dt \int_C ds \left\{\theta(y-x)\frac{1-ys}{ys(1-s)}
u^{pl}
+ \theta(x-y)\frac{1-xs}{xs(1-s)}v^{pl} \right\}
 \notag\\
&&\left.\quad
-\frac{(2lp+1)(lp+1)}{2\pi lp}\tan(lp\pi)
\int_0^1dt\int_0^1 ds w^{pl} \right] C^\nu_l{}'(0),
\label{Gn:IP}
}
where
\Eqrsub{
u &=& \frac{x}{y}\frac{t(1-t)}{1-tx}\frac{1-sy}{s(1-s)},\\
v &=& \frac{y}{x}\frac{t(1-t)}{1-ty}\frac{1-sx}{s(1-s)},\\
w &=& xy \frac{t(1-t)}{1-tx}\frac{s(1-s)}{1-sy}.
}
%

\subsection{Non-uniqueness for $n=2$}

Unfortunately, we have not succeeded in reducing the expression for $G_n(x,y)$ to a simple tractable form for $n>3$ yet. However, we can make such a reduction for $n=2$.

For $n=2$, the expression \eqref{Gn:IP} reads
\Eqr{
&& G_2(x,y) =  \frac{2}{(1-x)}\pfrac{x}{y}^{2}
\lim_{p\tend1}
\frac{1}{4\pi^2  (p-1)}
\notag\\
&&\quad\times
\int_0^1 dt \int_C ds \left\{\theta(y-x)\frac{1-ys}{ys(1-s)}
U_p
+ \theta(x-y)\frac{1-xs}{xs(1-s)}V_p \right\},
}
where
\Eq{
U_p = \sum_{l=2}^\infty \frac{2l+1}{2l}u^{pl} C_l^\nu{}'(0),\quad
V_p = \sum_{l=2}^\infty \frac{2l+1}{2l}v^{pl} C_l^\nu{}'(0).
}

$U_p$ and  $V_p$ can be calculated in the following way. First, utilising the generating function for $C^\nu_l(z)$, we obtain
\Eqr{
H(u,z) &:=& \sum_{l=2}^\infty \frac{2l+1}{2l} u^l C_l^\nu(z)
=\sum_{l=2}^\infty \insbra{u^l C_l^\nu(z)+\frac{1}{2}
\int_0^u du u^{l-1}C_l^\nu(z)}
\notag\\
&=& \frac{1}{(1-2zu+u^2)^{\nu}}-1+2\nu z u
\notag\\
&&\quad
+\frac{1}{2}\int_0^u\frac{du}{u}
\inrbra{\frac{1}{(1-2zu+u^2)^{\nu}}-1+2\nu z u}.
}
From this, for the present case with $\nu=1/2$, it follows that
\Eq{
U_p=u^p g(u^p);
\quad
g(x)=\frac{3}{2}-\frac{1}{2(1+x^2)^{1/2}}-\frac{1}{(1+x^2)^{3/2}}.
}
Hence, we have
\Eqr{
&& G_2(x,y) =  \frac{1}{2\pi^2(1-x)}\pfrac{x}{y}^{2}
\lim_{p\tend1}
\frac{1}{(p-1)}
\notag\\
&&\quad\times
\int_0^1 dt \int_C ds \left\{\theta(y-x)\frac{x}{y^2}
\frac{t(1-t)}{1-tx}\pfrac{1-sy}{s(1-s)}^{p+1} g(u^p)
\right.\notag\\
&&\left.\qquad
+\theta(x-y)\frac{y}{x^2}
\frac{t(1-t)}{1-ty}\pfrac{1-sx}{s(1-s)}^{p+1} g(v^p)\right\}.
}

Here, for $0<x\le1$, we can calculate the $p\tend1$ limit of the contour integral along $C$ as
\Eqr{
&&\frac{d}{dp}\int_C ds\pfrac{1-sy}{s(1-s)}^{p+1}g(u^p)
\notag\\
&&\quad
=\int_C ds \pfrac{1-sy}{s(1-s)}^2\inrbra{g(u)+ug'(u)}
 \ln\frac{1-sy}{s(1-s)}
\notag\\
&&\quad
=\frac{3}{2}\times 2\pi i \inpare{-\int_{C_0}+\int_{C_1}}ds \pfrac{1-sy}{s(1-s)}^2
\notag\\
&&\quad
=3\pi i \times 2\pi i \times (-2) 2(1-y) 
= 24\pi^2 (1-y),
}
where $C_0$ and $C_1$ are clockwise circle contours around $s=0$ and $s=1$, respectively. Similarly, for $0<y\le1$, we have
\Eq{
\frac{d}{dp}\int_C ds\pfrac{1-sx}{s(1-s)}^{p+1}g(v^p)
=24\pi^2 (1-x).
}
Hence, we find that $G_2(x,y)$ for $0<x,y\le1$ can be expressed as
\Eq{
G_2(x,y)=\frac{2}{1-x}\insbra{\theta(y-x) x^3 F_1(x)
\frac{F_2(y)}{y^4}
+\theta(x-y) F_2(x) \frac{F_1(y)}{y}}_{l=1},
}
where $l=1$ implies that $F_1$ and $F_2$ are those for $l=1$. This implies that if $s(x)$ vanishes around $x=0$, we have
\Eq{
\int_0^1 dy K_2(x,y)s(y)
=\LL_1(x)\int_0^1 dy G_2(x,y)s(y)=-\frac{6}{x^2(1-x)}s(x).
}
However, when $s(y)$ is not zero at $y=0$, the integral over $y$ on the left-hand side of this equation diverges and the equation becomes ill-defined. 

This difficulty can be resolved in the following way. First, because $s(y)-s(0)$ vanishes at $y=0$, the equation
\Eq{
\int_0^1 dy K_2(x,y)\inpare{s(y)-s(0)}
 = -\frac{6}{x^2(1-x)}\inpare{s(x)-s(0)}
}
should hold. Next, for the constant tension, $s(x)=2M\sigma(=\const)$, we know that the brane equation is satisfied. Hence, we have
\Eq{
\int_0^1 dy K_2(x,y) s(0)= 0.
}
These two equations together uniquely determine the action of the kernel $K_s(x,y)$ on a generic $s(y)$ as
\Eq{
\int_0^1 dy K_2(x,y)s(y)=\frac{6}{x^2(1-x)}\inpare{s(0)-s(x)}.
}
Inserting this result to \eqref{IPfors} with $n=2$, we find that the brane constraint reduces simply to
\Eq{
s(0)=2M\sigma.
}
This constrains only the value of $s(x)$ at $x=0$, i.e. the tension of the string at infinity, and does not restrict the $x$-dependence of $s(x)$. This implies that even if the additional constraints \eqref{n=2:s(x):constraint2} and \eqref{n=2:s(x):constraint2} coming from the asymptotic behaviour are taken into account, the brane constraint allows for infinitely many solutions each of which gives a localised braneworld black hole solution at least in the perturbative sense for the four-dimensional bulk case.

\section{Summary and Discussions}
\label{sec:summary}

In the present paper, we have constructed a static perturbative solution to the vacuum Einstein equations representing a black hole accelerated by a stringy source in higher dimensions. We have shown that such a solution always exists and is completely determined by a function $s(x)$ representing the local tension of the stringy source under the regularity condition at horizon and a natural asymptotic condition at infinity. We also pointed out that such a solution has no naked singularity but instead its horizon has a non-compact tubular structure extending to infinity in five or higher dimensions, unlike the four-dimensional C-metric. This feature is consistence with the uniqueness theorem for a higher-dimensional static black hole and the fact that the stringy singularity has a codimension equal to or greater than 3 in higher dimensions.

We then derived an integral equation for $s(x)$ that represents the condition for the existence of a hypersurface satisfying the vacuum brane condition. Each solution to this equation gives a localised static braneworld black hole solution in the perturbative sense, i.e., in the small mass limit. Hence, the existence and uniqueness of a solution to this constraint equation is closely related to the existence and uniqueness of a localised static braneworld black hole solution in the small mass limit. Unfortunately, due to the intricate structure of this constraint equation, we have not arrived at a complete answer to this problem, but we were able to obtain some interesting partial results. 

First, we have shown that there exists no hypersurface satisfying the vacuum brane condition if the string accelerating the black hole has a constant tension in higher dimensions, in contrast to the C-metric. This is a rather unexpected result because the non-uniformity of the string tension is equivalent to the condition that $T^t_t\neq T^r_r$. 

Second, we have found that there exist infinitely many localised static regular braneworld black hole solutions in the perturbative sense when the bulk spacetime is four dimensional. This is a quite embarrassing result from a classical point of view. However, it might be justified from the adS/CFT point of view, because in this point of view, a localised black hole solution on a brane is a solution to the quantum corrected field equations that might contain higher-derivative terms leading to the nonuniqueness of the solution\cite{Gregory.R&Ross&Zegers2008A}. For example, the fact that a non-trivial black hole solution exists on the 3-dimensional brane itself may be an evidence for that, because the vacuum Einstein equations allow only locally trivial solutions in three dimensions. 

Anyway, it will be quite important to check whether this nonuniqueness survives in the exact non-linear treatment. It is also a challenging task to extend the analysis to higher-dimensional cases.

\section*{Acknowledgements}

The author would like to thank Takahiro Tanaka and Simon Ross for valuable comments and the staff and members of CECS at Valdivia, Chile for their hospitality, where a part of this work was done. The author was supported in part by Grants-in-Aid for Scientific Research from JSPS (No. 18540265).
 
\appendix
\section{Spherical Harmonic Tensors}
\label{sec:HarmonicTensors}

In this appendix, we recapitulate the basic definitions and properties of the harmonic tensors on the $n$-dimensional unit Euclidean sphere $S^n$ with  the metric $ds^2= \gamma_{ij} dz^i dz^j$ and give explicit expressions for the scalar harmonics used in the present paper. We denote the covariant derivative with respect to $\gamma_{ij}$ by $\h D_i$. 

\subsection{General Definitions}

\subsubsection{Scalar harmonics}

Scalar harmonics, i.e., harmonic functions on a manifold $S^n$ are defined as eigenfunctions of the Laplace-Beltrami operator $\hat\triangle$ as
\Eq{
\hat\triangle \SHB=-k^2 \SHB.
}
The operator $\h\triangle$ is essentially self-adjoint in the function space $L^2(S^n)$ and has the discrete spectrum 
\Eq{
k^2= l(l+n-1),\quad l=0,1,\cdots.
}
The corresponding harmonic functions form a complete basis of $L^2(S^n)$. 

Here, note that the requirement on harmonic functions to belong to $L^2(S^n)$ is quite essential in determining the spectrum for $k^2$. In fact, for example, in the $\SO(n)$ symmetric case, the above eigenvalue problem can be written
\Eq{
\frac{1}{\sin^{n-1}\theta}\frac{d}{d\theta}\inpare{\sin^{n-1}\theta \frac{d}{d\theta} u(\theta)}=
-k^2 u(\chi)
\label{Harmonics:SO(n)-symmetric}}
and has always a solution for any value of $k^2$, if we do not impose any regularity condition on $u(\theta)$. The situations for vector and tensor harmonics are quite different as we discuss later.

From scalar harmonics, we can construct harmonic vectors by
\Eq{
\SHB_i= -\frac{1}{k} \hat D_i \SHB,
}
which satisfies
\Eq{
\triangle \SHB_i =-(k^2-n+1) \SHB_i.
}
Of course, this definition has meaning only for $k^2>0$.

Similarly, we can construct harmonic tensors by
\Eq{
\SHB_{ij}= \frac{1}{k^2} \h D_i \h D_j \SHB + \frac{1}{n}\gamma_{ij} \SHB,
}
which satisfies
\Eqrsub{
&& \triangle \SHB_{ij}=-(k^2-2n) \SHB_{ij},\\
&& \SHB^i_i=0.
}
This definition has meaning only for $k^2>0$ again.

\subsubsection{Vector harmonics}

Vector harmonics are vector fields on $S^n$ defined by the conditions
\Eqrsub{
&& \h\triangle \VHB_i = -k^2_v \VHB_i,\\
&& \h D_i \VHB^i=0.
}
When $\VHB_i$ is $L^2$-normalisable, the spectrum is given by
\Eq{
k^2_v= l(l+n-1)-1,\quad l=1,2,\cdots.
}
The corresponding harmonic vectors and the scalar-type harmonic vectors $\SHB_i$ form a complete basis of $L^2$-normalisable vector fields on $S^n$.

From vector harmonics, we can construct vector-type harmonic tensors by
\Eq{
\VHB_{ij}= -\frac{1}{2k_v} \inpare{ \h D_i \VHB_{j} + \h D_j \VHB_i },
}
which satisfy
\Eqrsub{
&& \h\triangle \VHB_{ij}= -(k^2_v-n-1) \VHB_{ij}.\\
&& \VHB^i_i=0.
}
Note that for the lowest eigenvalue $k_v^2=n-1$ (i.e., $l=1$), $\VHB_{ij}$ vanishes identically because the corresponding vector harmonic is a Killing vector.

\subsubsection{Tensor harmonics}

Finally, tensor harmonics are defined as 2nd-rank symmetric tensor fields on $S^n$ satisfying the conditions
\Eqrsub{
&& \h\triangle \THB_{ij} = -k_t^2 \THB_{ij},\\
&& \THB^i_i=0,\quad \h D_j \THB_i^j=0.
}
If we require the $L^2$-normalisability, the spectrum of $k_t^2$ for the Euclidean sphere is given by
\Eq{
k_t^2=l(l+n-1)-2,\quad l=2,3,\cdots.
}
The corresponding tensor harmonics together with harmonic tensors constructed from $\SHB$ and $\VHB_i$ form a complete basis for $L^2$-normalisable 2nd-rank symmetric tensor fields on $S^n$.

\subsection{Spherical Harmonics}

In this subsection, we recapitulate formulas for $\SO(n)$-symmetric harmonic functions $\SHB=Y_l$ on the Euclidean unit sphere $S^n$. In the coordinate system in which the metric is expressed as
\Eq{
ds^2= d\theta^2 + \sin^2\theta d\Omega_{n-1}^2,
}
we can assume that $Y_l$ depends only on $\theta$ and obeys the equation
\Eq{
\hat\triangle_n Y_l=\frac{1}{\sin^{n-1}\theta}
    \partial_\theta \left(\sin^{n-1}\theta\partial_\theta Y_l\right)
        =-l(l+n-1)Y_l.
}
The normalisable solution of this equation is given by
\Eq{
Y_l(\theta)= C^{(n-1)/2}_l(\cos\theta),
}
where $C^\nu_l(x)$ is the Gegenbauer polynomial normalised as
\Eqr{
&& \int_0^\pi d\theta\,\sin^{n-1}\theta C_l^{(n-1)/2}(\cos\theta)
       C_{l'}^{(n-1)/2}(\cos\theta)
       =\frac{8\pi\Gamma(l+n-1)}{2^n(2l+n-1)l![\Gamma(\frac{n-1}{2})]^2},\\
&& C_l^{(n-1)/2}(\pm1)=\frac{\Gamma(l+n-1)}{l!\Gamma(n-1)}(\pm1)^l.
}

The $\delta$ function on $S^n$ with support at the south pole, $\delta^n(-\Omega)$, can be expanded in terms of these harmonic functions as
\Eq{
\delta^n(-\Omega)=\sum_{l=0}^\infty a_l C_l^{(n-1)/2}(\cos\theta),
\label{deltabyCl}
}
where $a_l$ is determined from the above normalisation condition as
\Eq{
a_l=(-1)^l\frac{2^n(2l+n-1) [\Gamma(\frac{n-1}{2})]^2}
        {8\pi \Gamma(n-1)\Omega_{n-1}}.
\label{al}}
Here, $\Omega_{n}$ is the volume of the unit sphere $S^{n}$ and given by
\Eq{
\Omega_{n-1}=\frac{2\pi^{n/2}}{\Gamma(n/2)}.
}
%

\section{$\SO(n)$-Symmetric Tensors on $S^n$}
\label{sec:SOStensors}

In this appendix, we show that $\SO(n)$-symmetric perturbations of a $(n+2)$-dimensional Schwarzschild solution are of the scalar type if they are regular in directions corresponding to a hemisphere of the horizon.  For that purpose, we determine all possible vector and 2nd-rank symmetric tensor fields on $S^n$ that are $\SO(n)$-symmetric and of the vector or tensor type.

\subsection{Vectors}

Firstly, we consider a vector field $v_i$ on $S^n$. In general, it can be decomposed into the scalar and vector parts as
\Eq{
v_i = \h D_i s + v^{(1)}_i,\quad \h D_i v^{(1)i}=0.
}
In this section, we assume that $v_{i}$ and  $s$ are distributions on $S^n$ and the differentiation should be understood in the sense of distribution.

In the above decomposition, the scalar component $s$ satisfies 
\Eq{
\h \triangle s= \h D^i v_i.
}
A smooth function that is orthogonal to the left-hand side of this equation for any $s$ has to be a constant, and is always orthogonal to the right-hand side. Hence, this equation always has a solution that is unique up to the addition of a constant. Hence, the above decomposition of a vector to the scalar and the vector parts is always possible and effectively unique. 

Further, the scalar part $s$ has to be $\SO(n)$ invariant if $v_i$ is. Conversely, if $s$ is an $\SO(n)$-invariant function, $\h D_i s$ is an $\SO(n)$-invariant vector field. Therefore, we need to classify $\SO(n)$-invariant divergence-free vectors.

Here, note that in the coordinate system $(x^i)=(\theta, z^A)$ in which the metric of the Euclidean unit sphere $S^n$ is written 
\Eqr{
&& ds^2=d\theta^2 + \sin^2\theta d\Omega_{n-1}^2, 
\label{S^n:metric}\\
&& d\Omega_{n-1}^2= \gamma_{AB} dz^A dz^B,
}
$\SO(n)$ acts only on  the coordinates $z^A$ for $S^{n-1}$. In these coordinates, if $v^i$ is $\SO(n)$ invariant, $v^A=0$. Hence, the divergence-free condition reads
\Eq{
\pd_\theta\inpare{\sin^{n-1}\theta v^\theta}=0.
}
The general solution to this equation
\Eq{
v^\theta=\frac{C}{\sin^{n-1}\theta}
}
is always singular at the two poles of $S^n$, $\theta=0,\pi$.  This implies that any $\SO(n)$-invariant vector satisfying our regularity condition is of the scalar type.

\subsection{2nd-rank symmetric tensors}

Next, we consider a 2nd-rank trace-free symmetric tensor field $t_{ij}$ on $S^n$. We can restrict considerations to a tracefree tensor, which can be decomposed into the scalar, vector and tensor parts as
\Eq{
t_{ij}= t^{(0)}_{ij}+ t^{(1)}_{ij}+ t^{(2)}_{ij},
}
where the first part is the scalar part that can be written in terms of a scalar field $s$ as
\Eq{
t^{(0)}_{ij}= \h D_i \h D_j s -\frac{1}{n} \gamma_{ij}\h \triangle s,
}
and the second part is the vector part that can be written in terms of a divergence-free vector field $t_i$ as
\Eq{
t^{(1)}_{ij}= \h D_i t_j + \h D_j t_i;\quad
\h D_i t^i=0.
}
The last part is the transverse and trace-free part:
\Eq{
\h D_j t^{(2)}{}^j_i=0,\quad
t^{(2)}{}^i_i=0.
}
As in the case of vectors, we assume that $t_{ij}$ and related tensor fields such as $s$ and $t_i$ are distributions on $S^n$.

From these definitions, it immediately follows that
\Eqrsub{
&& \h D^i \h D^j t_{ij}= \frac{n-1}{n}\h\triangle(\h \triangle +n)s,
\label{sbytij}\\
&& \h D^j t_{ij}= \frac{n-1}{n}\h D_i(\h\triangle +n)s
   +(\h\triangle +n-1)t_i.
\label{tibytij}
}
A smooth function that is orthogonal to the right-hand side of \eqref{sbytij} for any distribution $s$ can be written as the sum of a constant and a harmonic function $Y$ corresponding to $l=1$. Here, the latter satisfies the differential relations\cite{Kodama.H2002a}
\Eq{
\h D_i \h D_j Y = -\gamma_{ij}Y,
\label{l1id}
}
from which it follows that $Y$ is orthogonal to the left-hand side of \eqref{sbytij}. Hence, \eqref{sbytij} can be always solved with respect to $s$. The solution is unique up to the addition of a constant and $Y$.

Similarly, smooth vector fields orthogonal to $(\h\triangle +n-1)t_i$ for any distributional vector field $t_i$ are spanned  by the Killing vectors of $S^n$, which are always orthogonal to the other terms in \eqref{tibytij}. Hence, \eqref{tibytij} can be always solved with respect to $t_i$, and the solution is unique up to the addition of a Killing vector.

Note that from these considerations it follows that if $t_{ij}$ is $\SO(n)$- symmetric, $s$ can be taken to be $\SO(n)$ symmetric as well, as is assumed in the present paper, because its $l=1$ part does not contribute to $t_{ij}$ owing to \eqref{l1id}. Note also that in the coordinates $(\theta,z^A)$ for $S^n$ introduced above, an $\SO(n)$-symmetric trace-free 2nd-rank symmetric tensor can be generally expressed as
\Eq{
t^i_j = [n-1,-1,\cdots,-1] f(\theta)= \hat t^i_j f(\theta),
\label{tij:generalform}
}
where $[v^1,\cdots,v^n]$ represents a diagonal matrix. Hence, the task is to determine all possible forms of $f(\theta)$.

The covariant derivative of this type of tensor $t^i_j$ is given by
\Eq{
\h D_k t^i_j =\h t^i_j\pd_k f 
+ n \insbra{\delta^\theta_j (\delta^i_k-\delta^i_\theta\delta^\theta_k)
+ \delta^i_\theta(g_{jk}-\delta^\theta_j\delta^\theta_k)}f \cot\theta.
}
In particular, the divergence of $t^i_j$ can be written
\Eq{
\h D_j t^j_i = \frac{1}{\sin^{n-1} \theta\sqrt{\gamma}}
   \pd_j\inpare{\sin^{n-1}\theta \sqrt{\gamma} f(\theta)}\h t^j_i
    + \Gamma^A_{Ai}f(\chi).
}
Further, the operation of the Laplacian is expressed as
\Eq{
\h\triangle_{n} t^i_j= \h t^i_j\insbra{ f'' + (n-1) f'\cot\theta 
   -2nf \cot^2\theta }.
}
%

\subsubsection{Tensor modes}

We first derive a condition for $t_{ij}$ to be divergence-free. This condition reduces to the single equation for $f(\theta)$, 
\Eq{
\h D_j t^j_\theta=\frac{n-1}{\sin^{n}\theta} 
   \pd_\theta \inpare{\sin^{n}\theta f(\theta)} =0.
}
The general solution to this equation is 
\Eq{
f(\theta)= \frac{C_T}{\sin^{n}\theta}.
\label{f:tensormode}
}
The corresponding tensor $t_{ij}$ is not $L^2$-normalisable and satisfies the harmonic equation with $k_t^2=-n$:
\Eq{
\h\triangle_{n} t^i_j= n t^i_j.
}
%

\subsubsection{Vector mode}

Next, we consider the vector-type solution that can be expressed as
\Eq{
t_{ij}=\h D_i t_j + \h D_j t_i,\quad \h D_i t^i=0.
}
Note that the divergence-free condition results from the first because $t_{ij}$ is trace free. 

First, from the equation for $t_{\theta\theta}$ 
\Eq{
2\pd_\theta t_\theta = (n-1) f(\theta)
}
$t_\theta$ is determined as
\Eq{
t_\theta= \frac{n-1}{2}F(\theta) + s(z);\quad F'=f.
}
Inserting this into the equation for $t_{\theta A}$, we obtain
\Eq{
  \sin^2\theta \pd_\theta\pfrac{t_A}{\sin^2\theta}
     + \pd_A s(z)
  = 0.
}
From this, we have
\Eq{
t_A= \sin\theta\cos\theta \pd_A s(z) + u_A(z) \sin^2\theta.
}
Finally, the equation for $t_{AB}$ reads
\Eqr{
 && 2\sin\theta\cos\theta D_A D_B s + (D_A u_B + D_B u_A)\sin^2\theta
       \notag\\
 && + 2\sin\theta\cos\theta \inpare{\frac{n-1}{2}F(\theta)+s(z)}
       \gamma_{AB}
       = -f(\theta)\sin^2\theta \gamma_{AB}.
}
This equation is equivalent to the following three equations:
\Eqrsub{
&& D_A D_B s = \frac{1}{n-1} D^2 s \gamma_{AB},\\
&& D_A u_B + D_B u_A =\frac{2}{n-1}(D\cdot u)\gamma_{AB},\\
&& (n-1)^2F(\theta)+2\inrbra{D^2 s + (n-1) s} + \inrbra{2D\cdot u+(n-1) f}\tan\theta.
}
The last of these is further equivalent to the following three equations
\Eqrsub{
&& D^2 s + (n-1) s= C_s,\\
&& D\cdot u = (n-1) C_u,\\
&& (n-1) F+ f\tan\theta + 2 C_u \tan\theta=0.
}
From now on, we set $C_s=0$ by shifting $s$ by a constant. Since the last equation is equivalent to 
\Eq{
f' + \inpare{(n-1)\cot\theta + \frac{1}{\sin\theta\cos\theta}} f 
 =-\frac{2C_u}{\sin\chi\cos\chi},
}
$f$ is determined as
\Eqr{
f &=& C f_{v1} + 2 C_u f_{v2};\label{f:vectormode}\\
f_{v1} &=& \frac{\cos\theta}{\sin^{n}\theta},\\
f_{v2} &=& -1+\frac{n-1\cos\theta}{\sin^{n}\theta}
        \int_0 \sin^{n-1}\theta d\theta.
}
The corresponding vector field $t_i$ can be written
\Eqrsub{
t_\theta &=& -\inpare{C_u + \frac{f}{2}} \tan\theta + s(z),\\
t_A &=& \inpare{ \pd_A s(z) \cos\theta + u_A(z) \sin\theta}\sin\theta.
}
Here, $s(z)$ and $u_A(z)$ are functions and vector fields on $S^{n-1}$ satisfying respectively
\Eqr{
&& D_A D_B s = - s\gamma_{AB},\\
&& D_A u_B + D_B u_A = 2 C_u \gamma_{AB}.
}
Solutions of the first equation are one-to-one correspondence with homogeneous coordinates of $S^{n-1}$, i.e., some Cartesian coordinate in the standard embedding of $S^{n-1}$ into $E^n$.  There exist $n$-independent such solutions. Next, for $C_u=0$, solutions to the second equation are in one-to-one correspondence with Killing vectors of $S^{n-1}$ and parametrised by $n(n-1)/2$ independent parameters. It is easy to see that these degrees of freedom altogether correspond to the freedom to add a Killing vector of $S^n$ to $t_i$ and do not affect the tensor field $t_{ij}$. Hence, we can set them to zero and assume that $t_i$ is also $\SO(n)$-symmetric, as is expected from the general argument at the beginning of this appendix. 

Now, we show that $C_u$ must be zero. First note that if there exists $u_A$ satisfying the above equation with $C_u\neq0$, then we can assume that it can be written as $u_A= D_A u$ for some function $u$ on $S^{n-1}$. Then, the equation for $u_A$ can be written
\Eq{
D_A D_B u=C_u \gamma_{AB}.
}
By applying $D^B$ to this equation, we obtain
\Eq{
D_A \triangle_{n-1} u + (n-2) D_A u=0.
}
Because we also have $\triangle_{n-1}u = (n-1) C_u$, this implies that $D_A u=0$, which contradicts the assumption $C_u\neq0$(cf. \citen{Kodama.H2002a}).

Note that the operation of the Laplacian on $t_{ij}$ corresponding to $f_{v1}$ is given by
\Eq{
\triangle_{n} t^i_j = 2 n t^i_j.
}

To summarise, if $t_{ij}$ expressed as \eqref{tij:generalform} is of the vector or tensor type, $f(\theta)$ should be given by either \eqref{f:vectormode} with $C_u=0$ or \eqref{f:tensormode}. Both of these, however, have singularities at the north pole and the south pole directions. Hence, they are not allowed if we require that perturbations are regular in all directions corresponding to a hemisphere.

\section{The behavior of $X$ and $Y$ at infinity}
\label{sec:XY:horizon:l>=2}

In this appendix, we determine the asymptotic behavior of $X$ and $Y$ for modes with $l\ge2$ at infinity. 

First, note that $X^{(l)}$ and $Y^{(l)}$ can be written
\Eqrsubl{XYbyhY}{
X^{(l)} &=& Y^{(l)}-2(1-x)\frac{d}{dx}Y^{(l)}
  =-Y^{(l)} -2\frac{d}{dx}\h Y^{(l)}.\\
Y^{(l)} &=& \frac{\h Y^{(l)}}{1-x},\\
\h Y^{(l)} &=& \frac{2(-1)^l}{n-1}
 \insbra{-S_2(x)F_1(x)-\h S_1(x) F_2(x) + A x^{1+p(l+1)}F_1(x)},
}
where
\Eqrsub{
S_2(x) &:=& x^{1+p(l+1)}\int_x^1 dy y^{-2-p(l+1)}s(y)F_2(y),\\
\h S_1(x) &:=& x^{-p(l-1)}\int_0^x dy y^{-1+p(l-1)}s(y) F_1(y).
}
Near $x=0$, $S_1(x)$ can be rewritten with the help of  partial integrations as
\Eqr{
S_1(x) &=&
 \frac{s(x)F_1(x)}{p(l-1)}-\frac{x(s(x)F_1(x))'}{p(l-1)[1+p(l-1)]}
  \notag\\
  && +\frac{x^{-p(l-1)}}{p(l-1)[1+p(l-1)]}
   \int_0^x dy y^{1+p(l-1)}(s(y)F_1(y))''.
}
Similarly, $S_2(x)$ can be rewritten as
\Eqr{
S_2(x)&=& x^{1+p(l+1)}\int_x^a dy y^{-2-p(l+1)} s(y)F_2(y) + Cx^{1+p(l+1)}
 \notag\\
&& = -\frac{s(a)F_2(a)x^{1+p(l+1)}-s(x)F_2(x)}{1+p(l+1)}
- \frac{(sF_2)'(a)x^{1+p(l+1)}-x (s(x)F_2(x))'}{p(l+1)[1+p(l+1)]}
   \notag\\
&&\quad
  \cdots 
  -\frac{(sF_2)^{(k)}(a)x^{1+p(l+1)}-x^k(s(x)F_2(x))^{(k)}}
   {[1+p(l+1)]\cdots[1+p(l+1)-k]}
   +Cx^{1+p(l+1)}
 \notag\\
&&\qquad
  +\frac{x^{1+p(l+1)}}{[1+p(l+1)]\cdots[1+p(l+1)-k]}
   \int_x^a dy \frac{(s(y)F_2(y))^{(k+1)}}{y^{1+p(l+1)-k}},
}
where $k$ is an integer satisfying the condition $0< 1+p(l+1)-k\le1$. 
Inserting these expansions into \eqref{XYbyhY}, we obtain \eqref{XY:infinity:l>=2}.

In general, from the above expressions, it follows that when $s(x)$ is smooth at $x=0$, $Y^{(l)}$ for $l\neq k(n-1)-1$ ($k=1,2,\cdots)$) can be expressed in terms of a function $N(x)$ that is regular at $x=0$ as
\Eq{
Y^{(l)}=N(x)+ \h A x^{1+p(l+1)}F_1(x)+ \h B x^{-p(l-1)}F_2(x).
}
For the special values $l= k(n-1)-1$ ($k=1,2,\cdots$), this expression is modified as 
\Eq{
Y^{(l)}=N(x)+x^{k+1}L(x)\ln(x) + \h A x^{1+p(l+1)}F_1(x)+ \h B x^{-p(l-1)}F_2(x),
}
where $L(x)$ is another function that is regular at $x=0$. In either case, $Y^{(l)}$ is bounded at $x=0$ when $\h B=0$. In this case, $Y^{(l)}$ is of $C^1$ class at $x=0$ and 
\Eq{
Y^{(l)}{}'(0)=-\frac{(n-2)s(0)}{(l-1)(l+1)(l+n)(l+n-2)}
  -\frac{s'(0)}{(l+1)(l+n-2)}.
}
However, it is not of $C^2$ class in general for $n\ge3$, while it is of $C^3$ class for $n=2$ and of $C^2$ class for $n=3$. 

When $Y^{(l)}$ is of $C^2$ class, we can calculate its second derivative at $x=0$ as
\Eqr{
Y^{(l)}{}''(0) &=& \frac{2(2n-3)(n-2)^2s(0)}{(l-1)(l+1)(l+n)(l+n-2)(2n+l-3)(l+2-n)}
\notag\\
&& +\frac{2(n-2)(2n-3)s'(0)}{(l+1)(l+n-2)(2n+l-3)(l+2-n)}
 -\frac{s''(0)}{(2n+l-3)(l+2-n)}
}
by differentiating the master equation multiplied by $x^2(1-x)$ with respect to $x$ twice and setting $x=0$. In particular, for $n=3$, we obtain
\Eq{
Y^{(l)}{}''(0)=\frac{6s(0)}{(l-1)^2(l+1)^2(l+3)^2}
 +\frac{6s'(0)}{(l-1)(l+1)^2(l+3)}
 -\frac{s''(0)}{(l-1)(l+3)}.
\label{Y''(0):n=3}
}
%

\section{Estimation of the Metric Perturbation Variables}
\label{sec:estimation:modesum}

In this appendix, we evaluate the behavior of the mode sums $\b X(x,\theta)$ and $\b Y(x,\theta)$ at spatial infinity and at horizon.

\subsection{Values at $r = \infty$}

The values of $\b X$ and $\b Y$ at spatial infinity can be written
\Eqrsub{
\b Y(x=0,\theta)
 &=&\sum_{l=2}^\infty\frac{-2(-1)^l(2l+n-1)s(0)}{(n+l)(l-1)}C_l^\nu(z) \notag\\
 &=& -2s(0) \inrbra{\b C_{-1} (-z)+\b C_n(-z)},
 \label{Y:infinity:series}\\
\b X(x=0,\theta) &=& \sum_{l=2}^\infty 
  \frac{2(-1)^l(2l+n-1)}{(l+n-2)(l+1)}
 \insbra{2s'(0)- \frac{l^2+(n-1)l+2-n}{(n+l)(l-1)}s(0)}C^\nu_l(z)
 \notag\\
 &=&\b Y(x=0,\theta) 
   +2\insbra{s'(0) -(n-2)s(0)}\inrbra{\b C_{1}(-z)+\b C_{n-2}(-z) },
   \label{X:infinity:series}
}
where
\Eqr{
&& \nu=\frac{n-1}{2},\quad  z=\cos\theta.\\
&& \b C_p (z)=\sum_{l=2}^\infty \frac{1}{l+p}C_l^\nu(z).
}
%

Utilising the generating function of the Gegenbauer polynomials
\Eq{
(1-2t z + t^2)^{-\nu}=1+2\nu zt +\sum_{l=2}^\infty t^l C_l^\nu(z),
}
we obtain the following integral expressions for $\b C_{-1}$ and $\b C_n$:
\Eqr{
\b C_{-1}(z) &=& \int_0^1 dt\sum_{l=2}^\infty t^{l-2}C_l^\nu(z)
=\int_0^1 \frac{dt}{t^2}\left[(1-2tz+t^2)^{-\nu}
         -1-2\nu zt\right],\\
\b C_{n}(z)
&=& \int_0^1 dt\sum_{l=2}^\infty t^{2\nu+l} C_l^\nu(z) \notag\\
&=& \int_0^1 dt t^{2\nu}\left[(1-2tz+t^2)^{-\nu}-1-2\nu z t\right]
 \notag\\
&=& \frac{n-1}{n}\left(1-\frac{nz}{n+1}\right)
   +\int_1^\infty \frac{dt}{t^2}\left[(1-2tz+t^2)^{-\nu }-1\right].
}
Therefore, with the help of partial integrations, we have
\Eqr{
G_1(z)&:=&\b C_{-1}(z)+ \b C_{n}(z) \notag\\
 &=& \frac{n-1}{n}\left(1+\frac{n^2z}{n+1}\right) 
    -2\nu \int_0^\infty dt \frac{1}{(1-2tz+t^2)^{\nu +1}} \notag\\
 && +2\nu z\lim_{\epsilon\tend0}\left(
    \int_0^\infty dt\frac{t^{-1+\epsilon}}{(1-2tz+t^2)^{\nu +1}}
      -\frac{1}{\epsilon}\right).
}
Utilising the formula
\Eqr{
&&\int_0^\infty dt \frac{t^\nu}{(1-2tz+t^2)^\mu} =
  \frac{2^{\mu-1/2}\Gamma(\nu+1)\Gamma(2\mu-\nu-1)}
    {\Gamma(2\mu)(1-z)^{\mu-1/2}}  \notag\\
&&\qquad \times   F(\mu-\nu-1/2,-\mu+\nu+3/2,\mu+1/2;(1+z)/2),
}
that is valid for $\nu>-1$ and $\nu-2\mu<-1$, this integral expression can be written in terms of hypergeometric functions as
\Eqr{
-\frac{G_1(-z)}{n-1}
  &=&
   -\frac{1}{n}-\left[\psi(n+1)-\psi(1)-\frac{n}{n+1}\right]\cos\theta 
   \notag\\
&& +\frac{2^{n/2}}{n(1+\cos\theta)^{n/2}}
     F\left(\frac{n}{2},-\frac{n}{2}+1,\frac{n}{2}+1;
                      \frac{1-\cos\theta}{2}\right) \notag\\
&& +\frac{2^{n/2}\cos\theta}{(1+\cos\theta)^{n/2}}
     \partial_\epsilon F\left(\frac{n}{2}+1-\epsilon,-\frac{n}{2}+\epsilon,
       \frac{n}{2}+1;\frac{1-\cos\theta}{2}\right).
}
This expression can be further deformed to
\Eqr{
-\frac{G_1(-z)}{n-1}
 &=&
   -\left[\psi(n+1)-\psi(1)-\frac{n}{n+1}\right]\cos\theta 
   -\frac{1-\cos\theta)}{n}
   \notag\\
&& \quad 
  +\pfrac{2}{1+cos\theta}^{n/2-1}
  \partial_\epsilon F\left(\frac{n}{2}+1-\epsilon,
     -\frac{n}{2}+\epsilon,\frac{n}{2};\frac{1-\cos\theta}{2}\right).
}
with the helps of the standard formulae for hypergeometric functions
\Eqr{
&& (\alpha+1-\beta)(1-z)F(\alpha+1,\beta,\gamma;z)
    +(\gamma-\alpha-1)F(\alpha,\beta,\gamma;z) \notag\\
&&\qquad +(\beta-\gamma)F(\alpha+1,\beta-1,\gamma;z)=0
}
with $\gamma=n/2+1, \alpha=n/2-\epsilon, \beta=-n/2+1+\epsilon$ and 
\Eqrsub{
&& (\gamma-\beta-1) F(\alpha.\beta,\gamma;z)
  +\beta F(\alpha,\beta+1,\gamma;z)
   =(\gamma-1)F(\alpha,\beta,\gamma-1;z),\notag\\
 &&\\
&& \gamma[F(\alpha,\beta+1,\gamma;z)-F(\alpha,\beta,\gamma;z)]
 =\alpha z F(\alpha+1,\beta+1,\gamma+1;z)
}
with $\alpha=n/2+1-\epsilon,\beta=-n/2+\epsilon,\gamma=n/2+1$.

By similar calculations, we obtain 
\Eqrsub{
\b C_l(z)
   &=&\int_0^1 dt \sum_{l=2}^\infty t^l C_l^\nu (z) \notag\\
  &=&\int_0^1 dt (1-2zt+t^2)^{-\nu } -1 -\nu z, \\
\b C_{n-2}(z)
   &=& \int_0^1 dt \sum_{l=2}^\infty t^{l+n-3} C_l^\nu (z) \notag\\
   &=& \int_1^\infty dt (1-2zt+t^2)^{-\nu } -\frac{1}{n-2} -z,
}
and 
\Eqr{
\b C_1(z)+\b C_{n-2}(z)
  &=&-\frac{n-1}{n-2}-\frac{n+1}{2}z
    +\int_0^\infty dt (1-2zt+t^2)^{-\nu } \notag\\
 &=& -\frac{n-1}{n-2}-\frac{n+1}{2}z \notag\\
 && +\frac{2^{n-2}}{(n-2)(1-z)^{n/2-1}}
  F\left(\frac{n}{2}-1,-\frac{n}{2}+2,\frac{n}{2};\frac{1+z}{2}\right).
}
Inserting these expressions into \eqref{X:infinity:series} and \eqref{Y:infinity:series}, we obtain \eqref{X:infinity} and \eqref{Y:infinity}.

\subsection{Values at horizon}

At horizon, the values of $\b X$ and $\b Y$ coincide and are given by
\Eqr{
&& \b X(1,\theta)= \b Y(1,\theta)
 =-\frac{2}{(n-1)^2} \int_0^1 dx\,x^{-p-1} s(x)G(\cos\theta,x),\\
&& 
G(z,x)=\sum_{l=2}^\infty C_l^\nu(-z)\frac{(lp+1)\Gamma(lp+1)^2}{\Gamma(2lp+1)} x^{lp}F_1(x).
}
Here, with the helps of the integral expression for the hypergeometric function
\Eq{
F_1(x)=\frac{\Gamma(2lp+2)}{\Gamma(lp+1)^2}
 \int_0^1 \pfrac{t(1-t)}{1-tx}^{lp}dt,
}
and the generating function for $C_l^\nu$, $G(z,x)$ can be written as
\Eqr{
G &=& \int_0^1 dt\sum_{l=2}^\infty C_l^\nu(-z) (lp+1)(2lp+1) u^{l}
\notag\\
  &=& \int_0^1 dt \left[\frac{1-(1+4p)u^2+(2p+1)u^3 z + (2p-1)uz}{(1+2zu+u^2)^{\nu+2}}
\right.
\notag\\
&&\left.\quad
 -1 + \frac{n(n+1)}{n-1} zu \right], 
}
where $u$ is defined by
\Eq{
u:=\pfrac{xt(1-t)}{1-tx}^p
}
Here, note that $u$ as a function of $t$ has the range for each $x$, 
\Eq{
0\le u \le \frac{4x^2}{\sqrt{1+x}(\sqrt{1-x}+\sqrt{1+x})^3}
\le 1,
}
where the $u$ takes the maximum value at $t=\frac{x}{1+\sqrt{1-x^2}}$.

Although these expressions for $\b X(1,\theta)$ and $\b Y(1,\theta)$ are not so enlightening, we can confirm at least that the total metric perturbation is regular except at the south pole $\theta=\pi$, i.e., $z=-1$. 


\end{document}